\documentclass[copyright,creativecommons,review]{eptcs}
\providecommand{\event}{Linearity\&TLLA 2020} 

\usepackage{breakurl}             

\usepackage{amsmath}
\usepackage{amssymb}
\usepackage{amsfonts}
\usepackage{amsthm}
\usepackage{todonotes}
\usepackage{tikz}
\usepackage{tikz-cd}
\usepackage{xcolor}
\usepackage{mathpartir}
\usepackage{stmaryrd}
\usepackage{multirow}
\usepackage{caption}
\usepackage{subcaption}
\usepackage{enumitem}
\usepackage{xypic}
\usepackage{float}

\usepackage{thmtools}
\usepackage{thm-restate}

\usepackage{listings}
\lstdefinelanguage{Granule}{%
  mathescape=true,
  morecomment=[l]{--},
  moredelim=[s][\itshape]{`}{`},
  showspaces=false,
  aboveskip=0.5em,
  belowskip=0.5em,
  commentstyle=\itshape\color{black!60},
  basicstyle=\ttfamily\footnotesize,
  flexiblecolumns=true,
  columns=[l]flexible,
  columns=fullflexible,
  keepspaces=true,
  xleftmargin=2.5em,
  literate=%
  {<}{\textcolor{effectColor}{<}}1
  {>}{\textcolor{effectColor}{>}}1
  {[}{\textcolor{coeffectColor}{[}}1
  {]}{\textcolor{coeffectColor}{]}}1
  {[r' : R']}{[{\textcolor{coeffectColor}{r' : R'}}]}1
  {[([}{\textcolor{coeffectColor}{[[}}2  
  {])]}{\textcolor{coeffectColor}{]]}}2
  {forall}{$\forall$}1
  {Inf}{$\infty$}1
  {->}{$\rightarrow$}1
  {-o}{$\multimap$}1
  {=>}{$\Rightarrow$}1
  {<-}{\textcolor{effectColor}{$\leftarrow$}}1
  {/\\}{$\sqcap$}1
  {\\/}{$\sqcup$}1
  {<=}{$\leqslant$}1
  {>=}{$\geqslant$}1
  {\\}{$\lambda$}1
  {_1}{$\mathtt{_1}$}1
  {_2}{$\mathtt{_2}$}1
  {_3}{$\mathtt{_3}$}1
  {_4}{$\mathtt{_4}$}1
  {_L}{$\mathtt{_{L}}$}1
  {_LH}{$\mathtt{_{LH}}$}1
  {_Gr}{$\mathtt{_{Gr}}$}1
  {_p}{$\mathtt{_{p}}$}1
  {_q}{$\mathtt{_{q}}$}1
  {-o}{$\multimap$}1
  {\\times}{$\times$}1
  {--BLANK}{}1,
  keywordstyle = \color{blue!40!black},
  keywords = {data, type, let, in, case, of, if, then, else, where,  import, Type, Semiring},
  keywordstyle = [4]\bfseries\color{effectColor},
  keywords     = [4]{pure}
  numbers=left,
  numberstyle=\tiny\color{gray}
}

\lstnewenvironment{granule}
{\lstset{language=Granule}}{}

\newcommand{\granin}[1]{\lstinline[language=Granule]{#1}}

\definecolor{multiplicity}{rgb}{0,0.3,0.08}

\lstdefinelanguage{Haskell}{%
  mathescape=true,
  morecomment=[l]{--},
  comment=[l]{\{-},
  moredelim=[s][\itshape]{`}{`},
  showspaces=false,
  aboveskip=0.5em,
  belowskip=0.5em,
  commentstyle=\itshape\color{black!60},
  basicstyle=\ttfamily\footnotesize,
  flexiblecolumns=true,
  columns=[l]flexible,
  columns=fullflexible,
  keepspaces=true,
  xleftmargin=2.5em,
  literate=%
   {\%r}{\textcolor{multiplicity}{\%r}}1
   {\%'Many}{\textcolor{multiplicity}{\%'Many}}1
   {\%1}{\textcolor{multiplicity}{\%1}}1,
  keywordstyle = \color{blue!40!black},
  keywords = {data, type, let, in, case, of, if, then, else, where,
  import, class, instance},
  numbers=left,
  numberstyle=\tiny\color{gray}
}

\lstnewenvironment{haskell}
{\lstset{language=Haskell}}{}

\newcommand{\haskin}[1]{\lstinline[language=Haskell]{#1}}

\theoremstyle{plain}

\theoremstyle{definition}
\newtheorem{definition}{Definition}[section]

\newtheorem{example}{Example}[section]
\newtheorem{remark}{Remark}

\newcommand{\synCase}[2]{\mathbf{case}\ #1\ \mathbf{of}\ #2}
\newcommand{\synCaseBranch}[2]{#1 \rightarrow #2}

\newcommand{\synCaseOne}[3]
  {\synCase{#1}{\synCaseBranch{#2}{#3}}}

\newcommand{\synCaseTwo}[5]
  {\synCase{#1}{\begin{array}[t]{l}\synCaseBranch{#2}{#3};\\ \synCaseBranch{#4}{#5}\end{array}}}

\newcommand{\synCaseTwoShort}[5]
  {\synCase{#1}{\begin{array}[t]{l}\synCaseBranch{#2}{#3};\;\; \synCaseBranch{#4}{#5}\end{array}}}

\newcommand{\synLetRec}[3]
 {\mathbf{letrec}\ #1 = #2\ \mathbf{in}\ #3}

\definecolor{coeffectColor}{HTML}{0a40a1}
\newcommand{\coeff}[1]{\textcolor{coeffectColor}{#1}}

\newcommand{\tcF}{\mathsf{F}}

\newcommand{\deriv}[2]{\llbracket{#2}\rrbracket_{\mathsf{#1}}}

\newcommand{\grminip}{\textsc{GrMini}$^P$}

\newcommand{\GRANULEdrule}[4][]{{\displaystyle\frac{\begin{array}{l}#2\end{array}}{#3}\quad\GRANULEdrulename{#4}}}

\newcommand{\GRANULEpremise}[1]{ #1 \\}
\newenvironment{GRANULEdefnblock}[3][]{ \framebox{\mbox{#2}} \quad #3 \\[0pt]}{}

\newcommand{\GRANULEnt}[1]{\mathit{#1}}
\newcommand{\GRANULEmv}[1]{\mathit{#1}}

\newcommand{\GRANULEsym}[1]{#1}

\newcommand{\GRANULEdrulename}[1]{\textsc{#1}}

\IfFileExists{xcolor.sty}{
	\RequirePackage{xcolor}
}{
	\RequirePackage{color}
}

 \usepackage{stmaryrd}
 \definecolor{coeffectColor}{HTML}{0a40a1} 
 \definecolor{effectColor}{HTML}{D64800}

\newcommand{\GRANULEdruleEquivXXcaseGenName}[0]{\GRANULEdrulename{Equiv\_caseGen}}
\newcommand{\GRANULEdruleEquivXXcaseGen}[1]{\GRANULEdrule[#1]{%
\GRANULEpremise{    \Gamma  \vdash  \GRANULEnt{t}  :   \square_{\textcolor{coeffectColor}{ \GRANULEmv{r} } }  \GRANULEnt{A}    \qquad    \GRANULEmv{r}   \vdash \,  \GRANULEnt{p_{\GRANULEmv{i}}}  :  \GRANULEnt{A}  \, \rhd \,  \Delta_{\GRANULEmv{i}}     \qquad     \Delta_{\GRANULEmv{i}}  \vdash   \GRANULEnt{p_{\GRANULEmv{i}}}   :  \GRANULEnt{A}   \qquad    \textcolor{coeffectColor}{1}   \, \textcolor{coeffectColor}{\sqsubseteq} \,  \GRANULEmv{r}    }%
}{
 \Gamma  \vdash   \textbf{case} \  \GRANULEnt{t}  \ \textbf{of} \   \overline{  \GRANULEsym{[}  \GRANULEnt{p_{\GRANULEmv{i}}}  \GRANULEsym{]}  \rightarrow   \GRANULEnt{p_{\GRANULEmv{i}}}   }    \equiv   \textbf{case} \  \GRANULEnt{t}  \ \textbf{of} \  \GRANULEsym{[}  \GRANULEmv{x}  \GRANULEsym{]}  \rightarrow  \GRANULEmv{x}   :  \GRANULEnt{A} }{%
{\GRANULEdruleEquivXXcaseGenName}{}%
}}

\newcommand{\GRANULEdruleEquivXXcaseAssocName}[0]{\GRANULEdrulename{Equiv\_caseAssoc}}

\newcommand{\GRANULEdruleEquivXXcaseBoxAssocName}[0]{\GRANULEdrulename{Equiv\_caseBoxAssoc}}

\renewcommand{\GRANULEdruleEquivXXcaseGenName}{\small{\textsc{CaseGen}}}

\renewcommand{\GRANULEdruleEquivXXcaseBoxAssocName}{\small{[\textsc{CaseAssoc}]}}
\renewcommand{\GRANULEdruleEquivXXcaseAssocName}{\small{\textsc{CaseAssoc}}}

\newif\ifappendix
\appendixfalse

\title{Deriving Distributive Laws for Graded Linear Types\vspace{-0.25em}}
\def\titlerunning{Deriving distributive laws for graded linear types}
\def\authorrunning{J. Hughes, M. Vollmer, D. Orchard}

\author{Jack Hughes
  \institute{School of Computing, University of Kent}
\and
  Michael Vollmer
  \institute{School of Computing, University of Kent}
 \and Dominic Orchard
\institute{School of Computing, University of Kent}}

\newif\iflongversion
\longversiontrue

\begin{document}
\maketitle

\begin{abstract}
  The recent notion of graded modal types provides a framework for
  extending type theories with fine-grained data-flow reasoning.  The
  Granule language explores this idea in the context of linear types.
  In this practical setting, we observe that the presence of graded
  modal types can introduce an additional impediment when
  programming: when composing programs, it is often necessary to
  `distribute' data types over graded modalities, and vice versa. In
  this paper, we show how to automatically derive these distributive
  laws as combinators for programming. We discuss the implementation
  and use of this automated deriving procedure in Granule, providing
  easy access to these distributive combinators.  This work is also
  applicable to Linear Haskell (which retrofits Haskell with linear
  types via grading) and we apply our technique there to provide the
  same automatically derived combinators. Along the way, we discuss
  interesting considerations for pattern matching analysis via graded
  linear types. Lastly, we show how other useful structural combinators
  can also be automatically derived.
\end{abstract}

\section{Introduction}
\label{sec:intro}
%

Linear type systems capture and enforce the idea that some data travels a \emph{linear}
dataflow path through a program, being consumed exactly once. This is
enforced by disallowing the structural rules of weakening and
contraction on variables carrying such values. Non-linear dataflow is
then typically captured by a modality $!A$ characterising
values/variables on which all structural rules are
permitted~\cite{girard1987linear}. This binary characterisation
(linear \textit{versus} non-linear) is made more fine-grained in Bounded
Linear Logic (BLL) via a family of modalities $!_r A$ where
$r$ is a polynomial term in $\mathbb{N}$ capturing the maximum number of times $r$ that a
value $A$ can be used~\cite{girard1992bounded}. The proof rules of BLL
track upper-bounds on non-linear use via these indices.
Various works have generalised the indices of BLL's modalities to arbitrary
semirings~\cite{brunel2014core,ghica2014bounded,petricek2014coeffects}
(often referred to as \emph{coeffect} systems). These generalised systems provide a unified
way to capture various program properties relating to
dataflow via \emph{graded necessity modalities}
$\Box_r A$ where $r$ is drawn from a
semiring.  The functional programming
language Granule\footnote{\url{https://granule-project.github.io/}}
provides a vehicle for exploring graded modal types
(both graded necessity and graded possibility, which can model side
effects~\cite{DBLP:conf/popl/Katsumata14}) alongside
linear and indexed types (GADTs)~\cite{DBLP:journals/pacmpl/OrchardLE19}. There
are various similar systems in the recent literature:
Abel and Bernardy give an alternate graded modal
calculus~\cite{abel-barnardy-icfp2020} and Linear Haskell
retrofits linearity onto the Haskell type
system~\cite{linear-haskell} via a graded type system akin to the
coeffect calculus of Petricek et
al.~\cite{petricek2014coeffects}. The popularity of graded types
is growing and their practical implications are being explored in
research languages
(e.g. Granule), and popular
functional languages (e.g. Haskell~\cite{linear-haskell} and Idris 2~\cite{bradyidris}).

When programming with graded modal types, we have observed
there is often a need to `distribute' a graded modality over a type, and
vice versa, in order to compose programs. That is,
we may find ourselves in possession of a $\Box_r (\mathsf{F} \alpha)$
value (for some parametric data type $\mathsf{F}$) which needs
to be passed to a pre-existing function (of our own codebase or a library) which requires
a $\mathsf{F} (\Box_r \alpha)$ value, or perhaps vice versa. A \emph{distributive
law} (in the categorical sense, e.g.,~\cite{street1972formal})
provides a conversion from one to the other.
In this paper, we present a procedure to automatically synthesise these
distributive operators, applying a generic programming
methodology~\cite{hinze2000new} to compute these operations given the
base type (e.g., $\mathsf{F} \alpha$ in the above description). This
serves to ease the use of graded modal types in practice, removing
boilerplate code by automatically generating these `interfacing
functions' on-demand, for user-defined data types as well as
built-in types.

Throughout, we refer to distributive laws of the form
$\Box_r (\mathsf{F} \alpha) \rightarrow \mathsf{F} (\Box_r \alpha)$
as \emph{push} operations (as they `push' the graded modality inside
the type constructor $\mathsf{F}$), and dually
$\mathsf{F} (\Box_r \alpha) \rightarrow \Box_r (\mathsf{F} \alpha)$
as \emph{pull} operations (as they `pull' the graded modality outside
the type constructor $\mathsf{F}$).

The primary contributions of this paper are then:
\begin{itemize}[itemsep=0em,leftmargin=1em]
  \item an overview of the application of distributive laws
    in the context of graded modal types;
  \item an automatic procedure for calculating distributive laws from
    types and a formal analysis of their properties;
  \item a realisation of this approach in both Granule (embedded into
    the compiler) and Linear Haskell (expressed within the language itself, leveraging
    Haskell's advanced features);
  \item and derivations of related combinators for structural use
    of values in a graded linear setting.
\end{itemize}
Along the way, we also analyse choices made around the typed-analysis
of pattern matching in various recent works on graded modal types.
Through the comparison between Granule and Linear Haskell, we also
highlight ways in which Linear Haskell could be made more general
and flexible in the future.

Section~\ref{sec:calculus} defines a core calculus \grminip{} with
linear and graded modal types which provides an idealised,
simply-typed subset of Granule with which we develop the core
contribution.  Section~\ref{sec:push-pull} gives the
procedures for deriving \emph{push} and \emph{pull} operators for the core calculus,
and verifies that these are distributive laws of endofunctors over the
$\Box_r$ graded comonadic modality.
Section~\ref{sec:implementation} describes the details of how these
procedures are realised in the Granule language.
Section~\ref{sec:linhaskell} relates this work to Linear Haskell, and
demonstrates how the \emph{push} and \emph{pull} combinators for user-defined data types
may be automatically generated at compile-time using Template Haskell.
Section~\ref{sec:matching-and-consumption} gives a comparison of the
recent literature with regards the typed (graded) analysis of pattern
matching, which is germane to the typing and derivation of our
distributive laws.
Section~\ref{sec:other} covers how other structural combinators for
Granule and Linear Haskell may be derived. Finally,
Section~\ref{sec:conclusion} discusses more related and future work.

We start with an extended motivating example typifying the kind
of software engineering impedance problem that distributive laws
solve. We use Granule since it is the main vehicle for the developments
here, and introduce some of the key concepts of graded modal types (in
a linear context) along the way.

\subsection{Motivating Example}
\label{sec:motivating-example}

Consider the situation of projecting the first element of a pair. In Granule,
this first-projection is defined and typed as the following
polymorphic function (whose syntax is reminiscent of Haskell or ML):
\begin{granule}
fst : forall {a b : Type} . (a, b [0]) -> a
fst (x, [y]) = x
\end{granule}
Linearity is the default, so this represents a linear function applied
to linear values.\footnote{Granule uses $\rightarrow$ syntax rather than
$\multimap$ for linear functions for the sake of familiarity with
standard functional languages} However, the second component of the pair
has a \emph{graded modal type}, written \granin{b [0]}, which means that we can use
the value ``inside'' the graded modality $0$ times by first `unboxing'
this capability via the pattern match \granin{[y]} which allows
weakening to be applied in the body to discard \granin{y} of type \granin{b}.
In calculus of Section~\ref{sec:calculus}, we denote
`\granin{b [0]}' as the type $\Box_0 b$ (Granule's graded
modalities are written postfix with the `grade' inside the box).

The type for \granin{fst} is however somewhat restrictive: what if
we are trying to use such a function with a value (call it
\granin{myPair}) whose type is not
of the form \granin{(a, b [0])} but rather \granin{(a, b) [r]} for
some grading term \granin{r} which permits weakening? Such a situation
readily arises when we are composing functional code, say between
libraries or between a library and user code. In this situation,
\granin{fst myPair} is ill-typed. Instead, we could define a
different first projection function for use with \granin{myPair : (a,
  b) [r]} as:
\begin{granule}
fst' : forall {a b : Type, s : Semiring, r : s} . {0 <= r} => (a, b) [r] -> a
fst' [(x, y)] = x
\end{granule}
This implementation uses various language features of Granule to make
it as general as possible. Firstly, the function is polymorphic
in the grade \granin{r} and in the semiring \granin{s} of which \granin{r}
is an element. Next, a refinement constraint \granin{0 <= r} specifies
that by the preordering \granin{<=} associated with the semiring
\granin{s}, that \granin{0} is approximated by \granin{r}
(essentially, that \granin{r} permits weakening). The rest
of the type and implementation looks more familiar for computing
a first projection, but now the graded
modality is over the entire pair.

From a software engineering perspective, it is cumbersome to
create alternate versions of generic combinators every time we are
in a slightly different situation with regards the position of a
graded modality.  Fortunately, this is an example to which a
general \emph{distributive law} can be deployed. In this case,
we could define the following distributive law of graded modalities over
products, call it \granin{pushPair}:
\begin{granule}
pushPair : forall {a b : Type, s : Semiring, r : s} . (a, b) [r] -> (a [r], b [r])
pushPair [(x, y)] = ([x], [y])
\end{granule}
This `pushes' the graded modality \granin{r} into
the pair (via pattern matching on the modality and the pair inside it, and then
reintroducing the modality on the right hand side via \granin{[x]} and
\granin{[y]}), distributing the graded modality to each component.
Given this combinator, we can now apply \granin{fst (pushPair myPair)} to yield a value of
type \granin{a [r]}, on which we can then apply the Granule standard library
function \granin{extract}, defined:
\begin{granule}
  extract : forall {a : Type, s : Semiring, r : s} . {(1 : s) <= r} => a [r] -> a
  extract [x] = x
\end{granule}
 to get the original \granin{a} value we desired:
\begin{granule}
extract (fst (pushPair myPair)) : a
\end{granule}
The \granin{pushPair} function could be provided by the standard
library, and thus we have not had to write any specialised combinators
ourselves: we have applied supplied combinators to solve the problem.

Now imagine we have introduced some custom data type \granin{List}
on which we have a \emph{map} function:
\begin{granule}
data List a = Cons a (List a) | Nil

map : forall {a b : Type} . (a -> b) [0..Inf] -> List a -> List b
map [f] Nil = Nil;
map [f] (Cons x xs) = Cons (f x) (map [f] xs)
\end{granule}
Note that, via a graded modality, the type of \granin{map} specifies that the parameter
function, of type \granin{a -> b} is non-linear, used between
$0$ and $\infty$ times. Imagine now we have a value
\granin{myPairList : (List (a, b)) [r]} and we want to map first
projection over it. But \granin{fst} expects \granin{(a, b [0])}
and even with \granin{pushPair} we require \granin{(a, b) [r]}.
\emph{We need another distributive law}, this time of the graded modality
over the \granin{List} data type. Since \granin{List} was
user-defined, we now have to roll our own \granin{pushList} operation, and so we
are back to having to make specialised combinators for our
data types.

The crux of this paper is that such distributive laws can be
automatically calculated given the definition of a type. With
our Granule implementation of this approach (Section~\ref{sec:implementation}),
we can then solve this combination problem via the following
composition of combinators:
\begin{granule}
map (extract . fst . push @(,)) (push @List myPairList) : List a
\end{granule}
where the \granin{push} operations are written with their base type
via \granin{@} (a type application) and whose definitions and types
are automatically generated during type checking. Thus the
\granin{push} operation is a \textit{data-type generic
  function}~\cite{hinze2000new}. This generic function is defined
inductively over the structure of types, thus a programmer can introduce a new
user-defined algebraic data type and have the implementation of the generic
distributive law derived automatically.
This reduces both the
initial and future effort (e.g., if an ADT definition changes or new ADTs are
introduced).

Dual to the above, there are situations where a programmer
may wish to \emph{pull} a graded modality out of a structure. This is
possible with a dual distributive law, which could be written
by hand as:
\begin{granule}
pullPair : forall {a b : Type, s : Semiring, m n : s} . (a [n], b [m]) -> (a, b) [n /\ m]
pullPair ([x], [y]) = [(x, y)]
\end{granule}
Note that the resulting grade is defined by the greatest-lower bound
(meet) of \granin{n} and \granin{m}, if it exists as defined
by a preorder for semiring \granin{s}
 (that is, $\sqcap$ is not a total operation). This
allows some flexibility in the use of the \emph{pull} operation when
grades differ in different components but have a
greatest-lower bound which can be `pulled out'.
Our approach also allows such operations to be generically derived.

%


\section{Core Calculus, \grminip}
\label{sec:calculus}
We define here a simplified monomorphic subset of Granule, which we call
the \grminip{} calculus.\footnote{\grminip{} is akin to the
\textsc{GrMini} calculus given
by Orchard et al.~\cite{DBLP:journals/pacmpl/OrchardLE19} but we include here the
notions of data type constructors, eliminations, and pattern matching
since these are important to the development here.}
This calculus extends the linear $\lambda$-calculus with a
\emph{semiring graded
  necessity modality}~\cite{DBLP:journals/pacmpl/OrchardLE19} where for a
preordered semiring
 $(\coeff{\mathcal{R}}, \coeff{*}, \coeff{1}, \coeff{+}, \coeff{0},
 \coeff{\sqsubseteq})$ (that is, a semiring with a pre-order $\coeff{\sqsubseteq}$ where $\coeff{+}$
 and $\coeff{*}$ are monotonic with respect to $\coeff{\sqsubseteq}$) there is a family of types
$\{\square_{\textcolor{coeffectColor}{ \GRANULEmv{r} } }  \GRANULEnt{A}\}_{\coeff{\GRANULEmv{r}} \in \coeff{\mathcal{R}}}$. Granule
allows multiple graded modalities (and thus multiple grading semirings) to be used simultaneously within a program,
but we focus here on just one graded modality, parameterising the
language. We also include notions of data constructor and their
elimination via \textbf{case} expressions as a way to unify the handling
of regular type constructors.

The syntax of \grminip{} terms and types is given by:
\begin{align*}
  \hspace{-0.9em}
  \GRANULEnt{t} ::= & \;
  \hspace{0.2em}\GRANULEmv{x}
  \hspace{0.2em}\mid \hspace{0.2em}\GRANULEnt{t_{{\mathrm{1}}}} \, \GRANULEnt{t_{{\mathrm{2}}}}
  \hspace{0.2em}\mid \hspace{0.2em}\lambda  \GRANULEmv{x}  .  \GRANULEnt{t}
  \hspace{0.2em}\mid \hspace{0.2em}\GRANULEsym{[}  \GRANULEnt{t}  \GRANULEsym{]}
  \hspace{0.2em}\mid \hspace{0.2em}C \, \GRANULEnt{t_{{\mathrm{0}}}} \, ... \, \GRANULEnt{t_{\GRANULEmv{n}}}
  \hspace{0.2em}\mid \hspace{0.2em}\textbf{case} \  \GRANULEnt{t}  \ \textbf{of} \   \GRANULEnt{p_{{\mathrm{1}}}}  \mapsto  \GRANULEnt{t_{{\mathrm{1}}}} ; ..;  \GRANULEnt{p_{\GRANULEmv{n}}}  \mapsto  \GRANULEnt{t_{\GRANULEmv{n}}}
  \hspace{0.2em}\mid \hspace{0.2em}\textbf{letrec}\  \GRANULEmv{x} \ =\  \GRANULEnt{t_{{\mathrm{1}}}} \ \textbf{in}\  \GRANULEnt{t_{{\mathrm{2}}}}
{\small{\tag{terms}}}
\\
  \hspace{-0.9em}
  \GRANULEnt{p} ::= & \;
  \hspace{0.2em} \GRANULEmv{x}
  \hspace{0.2em} \mid \hspace{0.2em} \_
  \hspace{0.2em }\mid \hspace{0.2em} \GRANULEsym{[}  \GRANULEnt{p}  \GRANULEsym{]}
  \hspace{0.2em} \mid \hspace{0.2em} C \, \GRANULEnt{p_{{\mathrm{0}}}} \, ... \, \GRANULEnt{p_{\GRANULEmv{n}}}
{\small{\tag{patterns}}}
\\
  \hspace{-0.9em}
  \GRANULEnt{A}, \GRANULEnt{B} ::= & \;
  \hspace{0.2em}\GRANULEnt{A}  \multimap  \GRANULEnt{B}
  \hspace{0.2em}\mid \hspace{0.2em} \alpha
  \hspace{0.2em}\mid \hspace{0.2em} \GRANULEnt{A}  \, \otimes \,  \GRANULEnt{B}
  \hspace{0.2em}\mid \hspace{0.2em} \GRANULEnt{A}  \, \oplus \,  \GRANULEnt{B}
  \hspace{0.2em}\mid \hspace{0.2em} \mathbf{1}
  \hspace{0.2em} \mid \hspace{0.2em} \square_{\textcolor{coeffectColor}{ \GRANULEmv{r} } }  \GRANULEnt{A}
  \hspace{0.2em}\mid \hspace{0.2em} \mu {X} .  \GRANULEnt{A}
  \hspace{0.2em}\mid \hspace{0.2em} {X}
{\small{\tag{types}}} \\
\hspace{-0.9em}
  C ::= & \; \mathsf{unit} \mid \mathsf{inl} \mid \mathsf{inr}
                \mid \GRANULEsym{(}  \,  \GRANULEsym{,}  \,  \GRANULEsym{)}
{\small{\tag{data constructors}}}
\end{align*}
Terms of \grminip{} consist of those of the linear $\lambda$-calculus,
plus a \emph{promotion} construct $\GRANULEsym{[}  \GRANULEnt{t}  \GRANULEsym{]}$ for introducing a graded modality,
data constructor introduction $C \, \GRANULEnt{t_{{\mathrm{0}}}} \, ... \, \GRANULEnt{t_{\GRANULEmv{n}}}$ and elimination ($\textbf{case}$), as
well as recursive let bindings ($\mathbf{letrec}$).
We will give more insights into the syntax via their
typing rules. Types comprise linear function types, type
variables ($\alpha$), multiplicative products, additive sums, unit
types, graded modal types, recursive types, and recursion variables
($X$) (which must be guarded by a $\mu$ binding).
Type variables are used in our derivation procedure but are
treated here as constants by the core calculus, without
any rules relating to their binding or unification.

\subsection{Typing}
 Typing is via judgments of the form $\Gamma  \vdash  \GRANULEnt{t}  :  \GRANULEnt{A}$, assigning
a type $A$ to term $t$ under the context $\Gamma$.
Typing contexts $\Gamma$ contain linear or graded
assumptions given by the grammar:
$$
\Gamma ::= \emptyset \mid \Gamma  \GRANULEsym{,}   \GRANULEmv{x}  :  \GRANULEnt{A} \mid \Gamma  \GRANULEsym{,}   \GRANULEmv{y}  : [  \GRANULEnt{A}  ]_{\textcolor{coeffectColor}{  \GRANULEmv{r}  } }$$
where $x$ is a linear assumption and
where $y$ is an assumption
graded by $\coeff{\GRANULEmv{r}}$ drawn from the preordered semiring
(called a \emph{graded assumption}). This delineation of linear and graded assumptions avoids issues with substitution under a promotion, following the technique of Terui's \textit{discharged assumptions}~\cite{Terui01lics}. Linear assumptions (including those of a graded modal type $\Box_{r} A$) must therefore be used exactly once, whilst graded assumptions may be used non-linearly subject to the constraint of $r$.

\begin{figure}[t]
    \begin{align*}
      \begin{array}{c}
\inferrule*[right=var]
{\quad}
{\GRANULEmv{x}  :  \GRANULEnt{A}   \vdash  \GRANULEmv{x}  :  \GRANULEnt{A}}
\qquad
\inferrule*[right=abs]
{\Gamma  \GRANULEsym{,}   \GRANULEmv{x}  :  \GRANULEnt{A}   \vdash  \GRANULEnt{t}  :  \GRANULEnt{B}}
{\Gamma  \vdash   \lambda  \GRANULEmv{x}  .  \GRANULEnt{t}   :   \GRANULEnt{A}  \multimap  \GRANULEnt{B}}
\qquad
\inferrule*[right=app]
{\Gamma_{{\mathrm{1}}}  \vdash  \GRANULEnt{t_{{\mathrm{1}}}}  :   \GRANULEnt{A}  \multimap  \GRANULEnt{B} \\ \Gamma_{{\mathrm{2}}}  \vdash  \GRANULEnt{t_{{\mathrm{2}}}}  :  \GRANULEnt{A}}
{\Gamma_{{\mathrm{1}}}  \GRANULEsym{+}  \Gamma_{{\mathrm{2}}}  \vdash  \GRANULEnt{t_{{\mathrm{1}}}} \, \GRANULEnt{t_{{\mathrm{2}}}}  :  \GRANULEnt{B}}
\\[0.7em]
\inferrule*[right=der]
{\Gamma  \GRANULEsym{,}   \GRANULEmv{x}  :  \GRANULEnt{A}   \vdash  \GRANULEnt{t}  :  \GRANULEnt{B}}
{\Gamma  \GRANULEsym{,}   \GRANULEmv{x}  : [  \GRANULEnt{A}  ]_{\textcolor{coeffectColor}{   \textcolor{coeffectColor}{1}   } }   \vdash  \GRANULEnt{t}  :  \GRANULEnt{B}}
\qquad
\inferrule*[right=weak]
{\Gamma  \vdash  \GRANULEnt{t}  :  \GRANULEnt{A}}
{\Gamma  \GRANULEsym{,}   [  \Delta  ]_{  \textcolor{coeffectColor}{0}  }   \vdash  \GRANULEnt{t}  :  \GRANULEnt{A}}
\qquad
\inferrule*[right=pr]
{\GRANULEsym{[}  \Gamma  \GRANULEsym{]}  \vdash  \GRANULEnt{t}  :  \GRANULEnt{A}}
{\textcolor{coeffectColor}{ \GRANULEmv{r}   \textcolor{coeffectColor}{\,*\,} }  \GRANULEsym{[}  \Gamma  \GRANULEsym{]}   \vdash  \GRANULEsym{[}  \GRANULEnt{t}  \GRANULEsym{]}  :   \square_{\textcolor{coeffectColor}{ \GRANULEmv{r} } }  \GRANULEnt{A}}
  \end{array}
\\[0.6em]
  \begin{array}{ccc}
    \inferrule*[right=approx]{\Gamma  \GRANULEsym{,}   \GRANULEmv{x}  : [  \GRANULEnt{A}  ]_{\textcolor{coeffectColor}{  \GRANULEmv{r}  } }   \GRANULEsym{,}  \Gamma'  \vdash  \GRANULEnt{t}  :  \GRANULEnt{A} \\ \GRANULEmv{r} \sqsubseteq \GRANULEmv{s}}
    {\Gamma  \GRANULEsym{,}   \GRANULEmv{x}  : [  \GRANULEnt{A}  ]_{\textcolor{coeffectColor}{  \GRANULEmv{s}  } }   \GRANULEsym{,}  \Gamma'  \vdash  \GRANULEnt{t}  :  \GRANULEnt{A}}
    &&
    \inferrule*[right=letrec]{\Gamma  \GRANULEsym{,}   \GRANULEmv{x}  :  \GRANULEnt{A}   \vdash  \GRANULEnt{t_{{\mathrm{1}}}}  :  \GRANULEnt{A} \\ \Gamma'  \GRANULEsym{,}   \GRANULEmv{x}  :  \GRANULEnt{A}   \vdash  \GRANULEnt{t_{{\mathrm{2}}}}  :  \GRANULEnt{B} }
    { \Gamma  \GRANULEsym{+}  \Gamma'  \vdash   \textbf{letrec}\  \GRANULEmv{x} \ =\  \GRANULEnt{t_{{\mathrm{1}}}} \ \textbf{in}\  \GRANULEnt{t_{{\mathrm{2}}}}   :  \GRANULEnt{B}}
\\[0.7em]
    \inferrule*[right=con]{(C : \GRANULEnt{B_{{\mathrm{1}}}}  \multimap \!\ldots\!\multimap  \GRANULEnt{B_{\GRANULEmv{n}}}  \multimap  \GRANULEnt{A}) \in \textsc{D}}
              { \emptyset\ \vdash\ C : \GRANULEnt{B_{{\mathrm{1}}}}  \multimap \!\ldots\!\multimap  \GRANULEnt{B_{\GRANULEmv{n}}}  \multimap  \GRANULEnt{A}}
        &
          &
\inferrule*[right=case]{
\Gamma  \vdash  \GRANULEnt{t}  :  \GRANULEnt{A} \quad\;
\cdot   \vdash \,  \GRANULEnt{p_{\GRANULEmv{i}}}  :  \GRANULEnt{A}  \, \rhd \,  \Delta_{\GRANULEmv{i}} \quad\;
\Gamma'  \GRANULEsym{,}  \Delta_{\GRANULEmv{i}}  \vdash  \GRANULEnt{t_{\GRANULEmv{i}}}  :  \GRANULEnt{B}}
{\Gamma  \GRANULEsym{+}  \Gamma'  \vdash   \textbf{case} \  \GRANULEnt{t}  \ \textbf{of} \   \GRANULEnt{p_{{\mathrm{1}}}}  \mapsto  \GRANULEnt{t_{{\mathrm{1}}}} ; ..;  \GRANULEnt{p_{\GRANULEmv{n}}}  \mapsto  \GRANULEnt{t_{\GRANULEmv{n}}}    :  \GRANULEnt{B}}
    \end{array}
 \end{align*}
\vspace{-0.8em}
 \caption{Typing rules for \grminip{}}
 \label{fig:typing-rules}
\end{figure}

Figure~\ref{fig:typing-rules} gives the typing rules.
We briefly explain each rule in turn.

The \textsc{var}, \textsc{abs} and \textsc{app} rules follow the standard rules for
the linear $\lambda$-calculus, modulo the \emph{context addition} operation
$\Gamma  \GRANULEsym{+}  \Gamma'$ which is used in the rules any time the contexts of two subterms
need to be combined. Context addition acts as union on disjoint
linear and graded assumptions but is defined via semiring addition
when a graded assumption appears in both contexts:
\begin{align*}
\GRANULEsym{(}  \Gamma  \GRANULEsym{,}   \GRANULEmv{x}  : [  \GRANULEnt{A}  ]_{\textcolor{coeffectColor}{  \GRANULEmv{r}  } }   \GRANULEsym{)}  \GRANULEsym{+}  \GRANULEsym{(}  \Gamma'  \GRANULEsym{,}   \GRANULEmv{x}  : [  \GRANULEnt{A}  ]_{\textcolor{coeffectColor}{  \GRANULEmv{s}  } }   \GRANULEsym{)}
= \GRANULEsym{(}  \Gamma  \GRANULEsym{+}  \Gamma'  \GRANULEsym{)}  \GRANULEsym{,}   \GRANULEmv{x}  : [  \GRANULEnt{A}  ]_{\textcolor{coeffectColor}{   \GRANULEmv{r}  \GRANULEsym{+}  \GRANULEmv{s}   } }
\end{align*}
Context addition $+$ is undefined in the case of contexts whose linear
assumptions are not-disjoint, enforcing the lack of contraction on
linear assumptions.

The next rules relate to grading. Structural weakening is provided by
(\textsc{weak}) for assumptions graded by 0 and `dereliction'
(\textsc{der}) converts a linear assumption to
a graded assumption, graded by 1.
Graded modalities are introduced by `promotion' (\textsc{pr}), scaling
graded assumptions in $\Gamma$ via semiring multiplication defined on
contexts containing only graded assumptions as:
\begin{align*}
    \textcolor{coeffectColor}{ \GRANULEmv{r}   \textcolor{coeffectColor}{\,*\,} }   \emptyset = \emptyset \qquad\qquad \textcolor{coeffectColor}{ \GRANULEmv{r}   \textcolor{coeffectColor}{\,*\,} }  \GRANULEsym{(}  \Gamma  \GRANULEsym{,}   \GRANULEmv{x}  : [  \GRANULEnt{A}  ]_{\textcolor{coeffectColor}{  \GRANULEmv{s}  } }   \GRANULEsym{)} = \GRANULEsym{(}   \textcolor{coeffectColor}{ \GRANULEmv{r}   \textcolor{coeffectColor}{\,*\,} }  \Gamma   \GRANULEsym{)}  \GRANULEsym{,}   \GRANULEmv{x}  : [  \GRANULEnt{A}  ]_{\textcolor{coeffectColor}{  \GRANULEsym{(}  \GRANULEmv{r}  \textcolor{coeffectColor}{\,*\,}  \GRANULEmv{s}  \GRANULEsym{)}  } }
\end{align*}
In the  (\textsc{pr}) rule, $\GRANULEsym{[}  \Gamma  \GRANULEsym{]}$ denotes contexts with only
graded assumptions.
The \textsc{approx} rule captures grade approximation, allowing a
grade $r$ in an assumption to be converted to another grade $s$, on the condition that
$r$ \textit{is approximated by} $s$, via the relation $\sqsubseteq$ as
provided by the semiring's preorder.

The \textsc{letrec} rule provides recursive bindings in the standard
way.

\noindent
Data constructors with zero or more arguments are introduced via
the \textsc{con} rule. Here, the constructors that concern us are
units, products, and coproducts (sums), given by $D$,
a global set of data constructors with their types, defined:
\begin{align*}
D = \{\mathsf{unit} : \mathbf{1}\}
\cup \{(,) : \GRANULEnt{A}  \multimap    \GRANULEnt{B}  \multimap   \GRANULEnt{A}  \, \otimes \,  \GRANULEnt{B} \mid \forall \GRANULEnt{A}, \GRANULEnt{B} \}
\cup \{\mathsf{inl} : \GRANULEnt{A}  \multimap   \GRANULEnt{A}  \, \oplus \,  \GRANULEnt{B} \mid \forall \GRANULEnt{A}, \GRANULEnt{B} \}
\cup \{\mathsf{inr} : \GRANULEnt{B}  \multimap   \GRANULEnt{A}  \, \oplus \,  \GRANULEnt{B} \mid \forall \GRANULEnt{A}, \GRANULEnt{B} \}
\end{align*}
Constructors are eliminated by pattern matching via the
\textsc{case} rule.  Patterns $p$ are typed by judgments of the form
$?\textit{r}  \vdash \,  \GRANULEnt{p}  :  \GRANULEnt{A}  \, \rhd \,  \Delta$ meaning that a pattern $p$ has type
$A$ and produces a context of typed binders $\Delta$ (used, e.g., in
the typing of the case branches). The information to the left of
the turnstile denotes optional grade information arising from being in
an unboxing pattern and is syntactically defined as either:
\begin{align*}
?\textit{r}\ ::= \cdot \mid \GRANULEmv{r}
\tag{enclosing grade}
\end{align*}
 where $\cdot$ means the present pattern is not nested inside an
unboxing pattern and $\GRANULEmv{r}$ that the present pattern is nested
inside an unboxing pattern for a graded modality with grade $\GRANULEmv{r}$.

\begin{figure}[t]
\begin{align*}
\setlength{\arraycolsep}{0em}
\begin{array}{ccc}
\inferrule*[right=Pvar]
 {\quad}
 {\cdot   \vdash \,  \GRANULEmv{x}  :  \GRANULEnt{A}  \, \rhd \,   \GRANULEmv{x}  :  \GRANULEnt{A}}
&
\inferrule*[right=Pcon]
{\cdot   \vdash \,  \GRANULEnt{p_{\GRANULEmv{i}}}  :  \GRANULEnt{B_{\GRANULEmv{i}}}  \, \rhd \,  \Gamma_{\GRANULEmv{i}}
}
{\cdot   \vdash \,  C \, \GRANULEnt{p_{{\mathrm{1}}}} \, .. \, \GRANULEnt{p_{\GRANULEmv{n}}}  :  \GRANULEnt{A}  \, \rhd \,  \Gamma_{{\mathrm{1}}}  \GRANULEsym{,} \, .. \, \GRANULEsym{,}  \Gamma_{\GRANULEmv{n}}}
&
\inferrule*[right={Pbox}]
{\GRANULEmv{r}   \vdash \,  \GRANULEnt{p}  :  \GRANULEnt{A}  \, \rhd \,  \Gamma}
{\cdot   \vdash \,  \GRANULEsym{[}  \GRANULEnt{p}  \GRANULEsym{]}  :   \square_{\textcolor{coeffectColor}{ \GRANULEmv{r} } }  \GRANULEnt{A}   \, \rhd \,  \Gamma}
\\[1em]
\inferrule*[right={[}Pvar{]}]
 {\quad}
 {\GRANULEmv{r}   \vdash \,  \GRANULEmv{x}  :  \GRANULEnt{A}  \, \rhd \,   \GRANULEmv{x}  : [  \GRANULEnt{A}  ]_{\textcolor{coeffectColor}{   \GRANULEmv{r}   } }}
\;\;&\;\;
\inferrule*[right={[}Pcon{]}]
{\GRANULEmv{r}   \vdash \,  \GRANULEnt{p_{\GRANULEmv{i}}}  :  \GRANULEnt{B_{\GRANULEmv{i}}}  \, \rhd \,  \Gamma_{\GRANULEmv{i}} \quad\quad
|  \GRANULEnt{A}  | > 1 \Rightarrow \textcolor{coeffectColor}{1}   \, \textcolor{coeffectColor}{\sqsubseteq} \,  \GRANULEmv{r}}
{\GRANULEmv{r}   \vdash \,  C \, \GRANULEnt{p_{{\mathrm{1}}}} \, .. \, \GRANULEnt{p_{\GRANULEmv{n}}}  :  \GRANULEnt{A}  \, \rhd \,  \Gamma_{{\mathrm{1}}}  \GRANULEsym{,} \, .. \, \GRANULEsym{,}  \Gamma_{\GRANULEmv{n}}}
\;\;&\;\;
\inferrule*[right={[}Pwild{]}]
 {\textcolor{coeffectColor}{0}   \, \textcolor{coeffectColor}{\sqsubseteq} \,  \GRANULEmv{r}}
 {\GRANULEmv{r}   \vdash \,   \_   :  \GRANULEnt{A}  \, \rhd \,   \emptyset}
\end{array}
\end{align*}
\caption{Pattern typing rules for \grminip{}}
\label{fig:pattern-rules}
\end{figure}

The rules of pattern typing are given in
Figure~\ref{fig:pattern-rules}.
The rule (\textsc{PBox}) provides
graded modal elimination (an `unboxing' pattern),
propagating grade information into the typing of the
sub-pattern. Thus $\textbf{case} \  \GRANULEnt{t}  \ \textbf{of} \  \GRANULEsym{[}  \GRANULEnt{p}  \GRANULEsym{]}  \rightarrow  \GRANULEnt{t'}$ can be used to eliminate
a graded modal value. Variable patterns are typed via two
rules depending on whether the variable occurs inside an unbox pattern
(\textsc{[Pvar]}) or not (\textsc{Pvar}),
with the \textsc{[Pvar]} rule producing a binding with the grade of
the enclosing box’s grade $\GRANULEmv{r}$.
As with variable patterns, constructor patterns are split
between rules for patterns which either occur inside an unboxing
pattern or not. In the former case, the grade information is
propagated to the subpattern(s), with the additional constraint that
if there is more than one data constructor for the type $\GRANULEnt{A}$ (written
$|A| > 1$), then the grade $r$ must approximate 1 (written $\textcolor{coeffectColor}{1}   \, \textcolor{coeffectColor}{\sqsubseteq} \,  \GRANULEmv{r}$) as pattern matching
incurs a usage to inspect the constructor (discussed further in
Section~\ref{subsec:matching-and-consumption}). The
operation $|  \GRANULEnt{A}  |$ counts the number of data constructors
for a type:
\begin{align*}
|   \mathbf{1}   | = 1 \;\;\;\;
|   \GRANULEnt{A}  \multimap  \GRANULEnt{B}   | = 1 \;\;\;\,
|   \square_{\textcolor{coeffectColor}{ \GRANULEmv{r} } }  \GRANULEnt{A}   | = |  \GRANULEnt{A}  | \;\;\;\;
|   \GRANULEnt{A}  \, \oplus \,  \GRANULEnt{B}   | = 2 (|A| + |B|) \;\;\;\;
|   \GRANULEnt{A}  \, \otimes \,  \GRANULEnt{B}   | = |A| |B| \;\;\;\;
|   \mu {X} .  \GRANULEnt{A}   | = |A [ \mu {X} .  \GRANULEnt{A} / X ]|
\end{align*}
and $|   {X}   |$ is undefined (or effectively 0) since we do not
allow unguarded recursion variables in types.
A type $A$ must therefore involve a sum type for $|A| > 1$.

Since a wildcard pattern $\_\,$ discards a value, this is only
allowed inside an unboxing pattern where the enclosing grade
permits weakening, captured via $\textcolor{coeffectColor}{0}   \, \textcolor{coeffectColor}{\sqsubseteq} \,  \GRANULEmv{r}$ in rule \textsc{[Pwild]}.


We now provide two examples of graded modalities. In addition to those
shown here, Granule provides
various other graded modalities, including grading by semirings of
security-level lattices
and ``sensitivities'' akin to \emph{DFuzz}~\cite{gaboardi2013linear}.
\begin{example}
  The natural numbers semiring $(\mathbb{N}, \ast, 1, +, 0, \equiv)$
  provides \emph{exact usage} where the preorder is equality
  $\equiv$. Graded modalities in this semiring count the number of
  times a value is used. As a simple example, we can define a function
  which copies its input value to produce a pair of values:
  \begin{align*}
\mathit{copy} & : \square_{\textcolor{coeffectColor}{ \GRANULEsym{2} } }  \GRANULEnt{A}    \multimap  \GRANULEsym{(}   \GRANULEnt{A}  \, \otimes \,  \GRANULEnt{A}   \GRANULEsym{)} \\
\mathit{copy} & = \lambda  \GRANULEmv{y}  .   \textbf{case} \  \GRANULEmv{y}  \ \textbf{of} \  \GRANULEsym{[}  \GRANULEmv{x}  \GRANULEsym{]}  \rightarrow   ( \GRANULEmv{x}  ,  \GRANULEmv{x} )
  \end{align*}
 The capability to use a value of type $\GRANULEnt{A}$ twice is
 captured by the graded modality. In the body of the abstraction,
variable $y : \square_{\textcolor{coeffectColor}{ \GRANULEsym{2} } }  \GRANULEnt{A}$ must be used linearly; $y$
is used linearly in the \textbf{case} which eliminates the
graded modality to yield a graded assumption $\GRANULEmv{x}  : [  \GRANULEnt{A}  ]_{\textcolor{coeffectColor}{  \GRANULEsym{2}  } }$ which can
be used twice to form the resulting pair.
\end{example}

\begin{example}
  BLL-style grading with upper bounds on usages can be obtained by
  replacing the equality relation in the $\mathbb{N}$-semiring with a
  less-than-equal ordering, giving a notion of
  \textit{approximation}. This is useful for programs that use
  variables differently in control-flow
  branches, e.g., when eliminating a sum type. This notion is
  further refined by the semiring of natural number intervals, which
  captures lower and upper bounds on usages. Intervals are given as
  pairs $\mathbb{N} \times \mathbb{N}$ written ${ \GRANULEmv{r} }..{ \GRANULEmv{s} }$, where $\GRANULEmv{r}$ and $\GRANULEmv{s}$ represent the lower and upper bounds respectively, with $\GRANULEmv{r} \leq \GRANULEmv{s}$, $0 =  {  \textcolor{coeffectColor}{0}  }..{  \textcolor{coeffectColor}{0}  }$ and $1 = {  \textcolor{coeffectColor}{1}  }..{  \textcolor{coeffectColor}{1}  }$. Ordering is given by $({ \GRANULEmv{r} }..{ \GRANULEmv{s} }) \leq ({ \GRANULEmv{r'} }..{ \GRANULEmv{s'} }) = \GRANULEmv{r'} \leq \GRANULEmv{r}
\wedge \GRANULEmv{s} \leq \GRANULEmv{s'}$. Using this, we can define the following
function to eliminate a coproduct via a pair of functions each of
which is used either 0 or 1 times:
  \begin{align*}
\oplus_e & :  \square_{[ {  \textcolor{coeffectColor}{0}  }..{  \textcolor{coeffectColor}{1}  } ]}(A \multimap C) \multimap  \square_{[ {  \textcolor{coeffectColor}{0}  }..{  \textcolor{coeffectColor}{1}  } ]}(B \multimap C) \multimap (\GRANULEnt{A}  \, \oplus \,  \GRANULEnt{B}) \multimap C\\
\oplus_e & =
 \lambda  \GRANULEmv{f'}  .    \lambda  \GRANULEmv{g'}  .   \lambda  \GRANULEmv{z}  .   \textbf{case} \  \GRANULEmv{f'}  \ \textbf{of} \  \GRANULEsym{[}  \GRANULEmv{f}  \GRANULEsym{]}  \rightarrow   \textbf{case} \  \GRANULEmv{g'}  \ \textbf{of} \  \GRANULEsym{[}  \GRANULEmv{g}  \GRANULEsym{]}  \rightarrow    \textbf{case} \  \GRANULEmv{z}  \ \textbf{of} \    \mathsf{inl}\  \GRANULEmv{x}   \mapsto   \GRANULEmv{f} \, \GRANULEmv{x}   ;   \mathsf{inr}\  \GRANULEmv{y}   \rightarrow  \GRANULEmv{g} \, \GRANULEmv{y}
  \end{align*}
\end{example}

\subsection{Equational Theory}

Figure~\ref{fig:equational} defines an equational theory for
\grminip{}. The equational theory is typed, but for brevity we elide
the typing for most cases since it follows
exactly from the structure of the terms. The fully typed
 rules are provided in the appendix~\cite{appendix}. For
those rules which are type restricted, we include the full
typed-equality judgment here. The typability of these equations relies on previous work on the Granule language which proves that pattern matching and substitution are well typed~\cite{DBLP:journals/pacmpl/OrchardLE19}.

\begin{figure}[h]
  \begin{align*}
    \begin{array}{c}
      \setlength{\arraycolsep}{0.15em}
  \begin{array}{rlr}
    \GRANULEsym{(}   \lambda  \GRANULEmv{x}  .  \GRANULEnt{t_{{\mathrm{2}}}}   \GRANULEsym{)} \, \GRANULEnt{t_{{\mathrm{1}}}} \ &\equiv \ \GRANULEnt{t_{{\mathrm{2}}}} [ \GRANULEnt{t_{{\mathrm{1}}}} / \GRANULEmv{x} ]  \ & \ \beta \\
    \lambda  \GRANULEmv{x}  .  \GRANULEnt{t}  \, \GRANULEmv{x} \ &\equiv \ \GRANULEnt{t}   \ & ( \GRANULEmv{x} \# \GRANULEnt{t} ) \
                                            \eta \\ \\
    \textbf{letrec}\  \GRANULEmv{x} \ =\  \GRANULEnt{t_{{\mathrm{1}}}} \ \textbf{in}\  \GRANULEnt{t_{{\mathrm{2}}}} &\equiv \ \GRANULEnt{t_{{\mathrm{2}}}} [ \textbf{letrec}\  \GRANULEmv{x} \ =\  \GRANULEnt{t_{{\mathrm{1}}}} \ \textbf{in}\  \GRANULEnt{t_{{\mathrm{1}}}} / \GRANULEmv{x} ]  & \ \beta_{letrec} \\
    \GRANULEmv{f} \, \GRANULEsym{(}   \textbf{letrec}\  \GRANULEmv{x} \ =\  \GRANULEnt{t_{{\mathrm{1}}}} \ \textbf{in}\  \GRANULEnt{t_{{\mathrm{2}}}}   \GRANULEsym{)} &\equiv \ \textbf{letrec}\  \GRANULEmv{x} \ =\  \GRANULEnt{t_{{\mathrm{1}}}} \ \textbf{in}\  \GRANULEsym{(}  \GRANULEmv{f} \, \GRANULEnt{t_{{\mathrm{2}}}}  \GRANULEsym{)}  &
                                                                     {\small{\textsc{LetRecDsitrib}}} \\
    \\
    \textbf{case} \  \GRANULEnt{t}  \ \textbf{of} \   \overline{  \GRANULEnt{p_{\GRANULEmv{i}}}  \rightarrow  \GRANULEnt{t_{\GRANULEmv{i}}}  } \ &\equiv \ (t \rhd \GRANULEnt{p_{\GRANULEmv{j}}} )\GRANULEnt{t_{\GRANULEmv{j}}}  \ & \;\; (\textsf{minimal}(j)) \ \beta_{case} \\
    \textbf{case} \  \GRANULEnt{t_{{\mathrm{1}}}}  \ \textbf{of} \   \overline{  \GRANULEnt{p_{\GRANULEmv{i}}}  \rightarrow   \GRANULEnt{t_{{\mathrm{2}}}}  \GRANULEsym{[}  \GRANULEnt{p_{\GRANULEmv{i}}}  \GRANULEsym{/}  \GRANULEmv{z}  \GRANULEsym{]}   } \ &\equiv \ \GRANULEnt{t_{{\mathrm{2}}}}  \GRANULEsym{[}  \GRANULEnt{t_{{\mathrm{1}}}}  \GRANULEsym{/}  \GRANULEmv{z}  \GRANULEsym{]} & \ \eta_{case} \\
   \textbf{case} \  \GRANULEsym{(}   \textbf{case} \  \GRANULEnt{t}  \ \textbf{of} \   \overline{  \GRANULEnt{p_{\GRANULEmv{i}}}  \rightarrow  \GRANULEnt{t_{\GRANULEmv{i}}}  }    \GRANULEsym{)}  \ \textbf{of} \   \overline{  \GRANULEnt{p'_{\GRANULEmv{i}}}  \rightarrow  \GRANULEnt{t'_{\GRANULEmv{i}}}  } &\equiv\
   \textbf{case} \  \GRANULEnt{t}  \ \textbf{of} \   \overline{  \GRANULEnt{p_{\GRANULEmv{i}}}  \rightarrow  \GRANULEsym{(}   \textbf{case} \  \GRANULEnt{t_{\GRANULEmv{i}}}  \ \textbf{of} \   \overline{  \GRANULEnt{p'_{\GRANULEmv{i}}}  \rightarrow  \GRANULEnt{t'_{\GRANULEmv{i}}}  }    \GRANULEsym{)}  } &\ {\small{\GRANULEdruleEquivXXcaseAssocName{}}}\\
   \GRANULEmv{f} \, \GRANULEsym{(}   \textbf{case} \  \GRANULEnt{t}  \ \textbf{of} \   \overline{  \GRANULEnt{p_{\GRANULEmv{i}}}  \rightarrow  \GRANULEnt{t_{\GRANULEmv{i}}}  }    \GRANULEsym{)} \ &\equiv\ \textbf{case} \  \GRANULEnt{t}  \ \textbf{of} \   \overline{  \GRANULEnt{p_{\GRANULEmv{i}}}  \rightarrow  \GRANULEsym{(}  \GRANULEmv{f} \, \GRANULEnt{t_{\GRANULEmv{i}}}  \GRANULEsym{)}  } & \ {\small{\textsc{CaseDistrib}}} \\
   \textbf{case} \  \GRANULEsym{[}   \textbf{case} \  \GRANULEnt{t}  \ \textbf{of} \   \overline{  \GRANULEnt{p_{\GRANULEmv{i}}}  \rightarrow  \GRANULEnt{t_{\GRANULEmv{i}}}  }    \GRANULEsym{]}  \ \textbf{of} \   \overline{  \GRANULEsym{[}  \GRANULEnt{p'_{\GRANULEmv{i}}}  \GRANULEsym{]}  \rightarrow  \GRANULEnt{t'_{\GRANULEmv{i}}}  } &\equiv\
   \textbf{case} \  \GRANULEsym{[}  \GRANULEnt{t}  \GRANULEsym{]}  \ \textbf{of} \   \overline{  \GRANULEsym{[}  \GRANULEnt{p_{\GRANULEmv{i}}}  \GRANULEsym{]}  \rightarrow    \textbf{case} \  \GRANULEsym{[}  \GRANULEnt{t_{\GRANULEmv{i}}}  \GRANULEsym{]}  \ \textbf{of} \   \overline{  \GRANULEsym{[}  \GRANULEnt{p'_{\GRANULEmv{i}}}  \GRANULEsym{]}  \rightarrow  \GRANULEnt{t'_{\GRANULEmv{i}}}  }     } & \;\; (\textsf{lin}( \GRANULEnt{p_{\GRANULEmv{i}}} )) \ {\small{\GRANULEdruleEquivXXcaseBoxAssocName{}}}\\
  \end{array}
\\ \\
\GRANULEdruleEquivXXcaseGen{}
    \end{array}
   \end{align*}
 \caption{Equational theory for \grminip{}}
 \label{fig:equational}
\end{figure}

The $\beta$ and $\eta$ rules follow the standard rules from the
$\lambda$-calculus, where $\#$ is a \textit{freshness} predicate,
denoting that variable $x$ does not appear inside term $t$.

For recursive \textbf{letrec} bindings, the $\beta_{letrec}$ rule substitutes
any occurrence of the bound variable $x$ in $\GRANULEnt{t_{{\mathrm{2}}}}$ with $\textbf{letrec}\  \GRANULEmv{x} \ =\  \GRANULEnt{t_{{\mathrm{1}}}} \ \textbf{in}\  \GRANULEnt{t_{{\mathrm{1}}}}$, ensuring that recursive uses of $x$ inside $\GRANULEnt{t_{{\mathrm{1}}}}$
can be substituted with $\GRANULEnt{t_{{\mathrm{1}}}}$ through subsequent $\beta_{letrec}$
reduction. The \textsc{LetRecDistrib} rule allows distributivity of
functions over \textbf{letrec} expressions, stating that if a function
$f$ can be applied to the entire \textbf{letrec} expression, then this
is equivalent to applying $f$ to just the body term $\GRANULEnt{t_{{\mathrm{2}}}}$.

Term elimination is via \textbf{case},
requiring rules for both $\beta$- and $\eta$-equality on
case expressions, as well as rules for associativity and
distributivity. In $\beta_{case}$, a term $t$ is matched against a
pattern $\GRANULEnt{p_{\GRANULEmv{j}}}$ in the context of the term $\GRANULEnt{t_{\GRANULEmv{j}}}$ through the use of the partial function
$(t\ \rhd\ \GRANULEnt{p_{\GRANULEmv{j}}})\GRANULEnt{t_{\GRANULEmv{j}}} = t'$ which may substitute terms bound in $\GRANULEnt{p_{\GRANULEmv{j}}}$
into $\GRANULEnt{t_{\GRANULEmv{j}}}$ to yield $t'$ if the match is successful. This partial function is defined inductively:
{{
\begin{align*}
\setlength{\arraycolsep}{0em}
\begin{array}{cccc}
\inferrule*[right=$\rhd_{-}$]
 {\quad}
  {( \GRANULEnt{t} \rhd \_) \GRANULEnt{t'} = \GRANULEnt{t'}}
\;
&
\inferrule*[right=$\rhd_{var}$]
 {\quad}
 {( \GRANULEnt{t} \rhd \GRANULEmv{x}) \GRANULEnt{t'} = \GRANULEsym{[}  \GRANULEnt{t}  \GRANULEsym{/}  \GRANULEmv{x}  \GRANULEsym{]}  \GRANULEnt{t'}}
\;
&
\inferrule*[right=$\rhd_{\Box}$]
 {( \GRANULEnt{t} \rhd \GRANULEnt{p}) \GRANULEnt{t'} = \GRANULEnt{t''}}
 {( \GRANULEsym{[}  \GRANULEnt{t}  \GRANULEsym{]} \rhd \GRANULEsym{[}  \GRANULEnt{p}  \GRANULEsym{]}) \GRANULEnt{t'} = \GRANULEnt{t''}}
\;
&
\inferrule*[right=$\rhd_{C}$]
 {( \GRANULEnt{t_{\GRANULEmv{i}}} \rhd \GRANULEnt{p_{\GRANULEmv{i}}}) \GRANULEnt{t}'_{i} = \GRANULEnt{t}'_{i+1}}
 {( C {t_{1}, .., t_{n}} \rhd C \, \GRANULEnt{p_{{\mathrm{1}}}} \, ... \, \GRANULEnt{p_{\GRANULEmv{n}}}) \GRANULEnt{t}'_{1} = \GRANULEnt{t}'_{n+1}}
\end{array}
\end{align*}
}}
As a predicate to the $\beta_{case}$ rule, we require that $j$ be
minimal, i.e.\ the first pattern $\GRANULEnt{p_{\GRANULEmv{j}}}$ in $\GRANULEnt{p_{{\mathrm{1}}}} ... \GRANULEnt{p_{\GRANULEmv{n}}}$ for
which $(t\ \rhd\ \GRANULEnt{p_{\GRANULEmv{j}}})\GRANULEnt{t_{\GRANULEmv{j}}} = t'$ is defined. Rule $\eta_{case}$ states that if all branches of the case expression share a common term $t_{2}$ which differs between branches only in the occurrences of terms that match the pattern used, then we can substitute $t_{1}$ for the pattern inside $t_{2}$.

Associativity of case expressions is provided by the
\textsc{CaseAssoc} rule. This rule allows us to restructure nested
case expressions such that the output terms $\GRANULEnt{t_{\GRANULEmv{i}}}$ of the inner
case may be matched against the patterns of the outer case, to achieve
the same resulting output terms $\GRANULEnt{t'_{\GRANULEmv{i}}}$. The \textsc{[CaseAssoc]}
rule provides a graded alternative to the \textsc{CaseAssoc} rule,
where the nested case expression is graded, provided that the patterns
$\GRANULEnt{p'_{\GRANULEmv{i}}}$ of the outer case expression are also graded. Notably,
this rule only holds when the patterns of the inner case expression
are linear (i.e., variable or constant) so that there are no nested box patterns, represented via the $\textsf{lin}( \GRANULEnt{p_{\GRANULEmv{i}}} )$ predicate. As with \textbf{letrec}, distributivity of functions over a case expression is given by \textsc{CaseDistrib}.

Lastly generalisation of an arbitrary boxed pattern to a variable is permitted through the \textsc{CaseGen} rule. Here, a boxed pattern $\GRANULEsym{[}  \GRANULEnt{p_{\GRANULEmv{i}}}  \GRANULEsym{]}$ and the output term of the case may be converted to a variable if the output term is equivalent to the pattern inside the box. The term $\GRANULEnt{t}$ being matched against must therefore have a grade approximatable by 1, as witnessed by the predicate $\textcolor{coeffectColor}{1}   \, \textcolor{coeffectColor}{\sqsubseteq} \,  \GRANULEmv{r}$ in the typing derivation.

\section{Automatically Deriving \emph{push} and \emph{pull}}
\label{sec:push-pull}
Now that we have established the core theory, we describe the
algorithmic calculation of distributive laws in \grminip{}.
Note that whilst \grminip{} is simply typed (monomorphic), it includes
type variables (ranged over by $\alpha$) to enable
the distributive laws to be derived on parametric types. In the implementation,
these will really be polymorphic type variables, but the derivation
procedure need only treat them as some additional syntactic type construct.


%
\paragraph{Notation}
Let $\tcF : \mathsf{Type}^n \rightarrow \mathsf{Type}$
be an $n$-ary type constructor (i.e. a constructor which takes $n$ type arguments), whose free type variables
provide the $n$ parameter types. We write $\tcF \overline{\alpha_i}$ for
the application of $\tcF$ to
type variables $\alpha_i$ for all $1 \leq i \leq n$.

%
\paragraph{Push}
We automatically
calculate \emph{push} for $\mathsf{F}$
applied to $n$ type variables
$\overline{\alpha_i}$
as the operation:
\begin{align*}
\deriv{push}{\tcF \overline{\alpha_i}} : \Box_{r} \tcF \overline{\alpha_i}
  \multimap  \tcF  (\overline{\Box_{r} \alpha_i})
 \end{align*}
where we require $\textcolor{coeffectColor}{1}   \, \textcolor{coeffectColor}{\sqsubseteq} \,  \GRANULEmv{r}\ \textit{if}\ |\tcF \overline{\alpha_i}| > 1$
due to the $[\textsc{Pcon}]$ rule (e.g., if $\tcF$ contains a sum).

For types $A$ closed with respect to recursion variables, let $\deriv{push}{A} = \lambda z
. \deriv{push}{A}^\emptyset\ z$ given by an intermediate
interpretation $\deriv{push}{A}^\Sigma$ where $\Sigma$ is a context of \textit{push} combinators for the
recursive type variables:
%
\begin{align*}
\deriv{push}{\mathbf{1}}^\Sigma \ z & = \textbf{case} \  \GRANULEmv{z}  \ \textbf{of} \  \GRANULEsym{[}    \mathsf{unit}    \GRANULEsym{]}  \rightarrow    \mathsf{unit}
\\
\deriv{push}{\alpha}^\Sigma       \ z & = z
  \\
  \deriv{push}{X}^\Sigma     \ z & = \Sigma(X)\ z
\\
\deriv{push}{A \oplus B}^\Sigma \ z & =
\synCaseTwoShort{z}{[\mathsf{inl}\ x]}{\mathsf{inl}\ \deriv{push}{A}^{\Sigma}[x]}{[\mathsf{inr}\ y]}
                            {\mathsf{inr}\
                                      \deriv{push}{B}^{\Sigma}[y]}
\\
\deriv{push}{A \otimes B}^\Sigma \ z & =
\synCaseOne{z}{[(x, y)]}
   {(\deriv{push}{A}^{\Sigma}[x], \deriv{push}{B}^{\Sigma}[y])}
\\
  \deriv{push}{A \multimap B} \ z & =
                                \lambda y .
                                \synCaseOne{z}{[f]}{\synCaseOne{\deriv{pull}{A}^{\Sigma}\
                                    y}{[u]}{\deriv{push}{B}^{\Sigma}[(f
                                \ u)]}}
\\
\deriv{push}{\mu X . A}^\Sigma \ z & =
 \synLetRec{f}{\deriv{push}{A}^{\Sigma, X \mapsto
f : \mu {X} .    \square_{\textcolor{coeffectColor}{ \GRANULEmv{r} } }  \GRANULEnt{A}      \multimap     \GRANULEsym{(}   \mu {X} .  \GRANULEnt{A}   \GRANULEsym{)}   \overrightarrow{ [    \square_{\textcolor{coeffectColor}{ \GRANULEmv{r} } }  \alpha_{\GRANULEmv{i}}    /  \alpha_{\GRANULEmv{i}}   ] } }}{f\ z}
\end{align*}
In the case of \emph{push} on a value of type $\mathbf{1}$, we pattern match on the value, eliminating the graded modality via the unboxing pattern match and returning the unit value. For type variables, \emph{push} is simply the identity of the value, while for recursion variables we lookup the $X$'s binding in $\Sigma$ and apply it to the value. For sum and product types, \emph{push} works by pattern matching on the type's constructor(s) and then inductively applying \emph{push} to the boxed arguments, re-applying them to the constructor(s).
Unlike \emph{pull} below, the \emph{push} operation can be derived for function
types, with a contravariant use of \emph{pull}.
For recursive types, we inductively apply \emph{push} to the value
with a fresh recursion variable bound in $\Sigma$, representing a
recursive application of push.

There is no derivation of a distributive law for types which are themselves graded
modalities (see further work discussion in Section~\ref{sec:conclusion}).

The appendix~\cite{appendix} proves that
$\deriv{push}{A}$ is type sound, i.e., its derivations are well-typed.

\paragraph{Pull}
We automatically
calculate \emph{pull} for $\mathsf{F}$
applied to $n$ type variables
$\overline{\alpha_i}$
as the operation:
\begin{align*}
\deriv{pull}{\tcF\ \overline{\alpha_i}} : \tcF\ (\overline{\Box_{r_i} \alpha_i})
\multimap \Box_{\bigwedge^n_{i = 1} r_i} (\tcF\  \overline{\alpha_i})
 \end{align*}
Type constructor $\tcF$ here is applied to $n$
arguments each of the form $\Box_{r_i} \alpha_i$, i.e., each with a
different grading of which the greatest-lower bound\footnote{The greatest-lower bound $\wedge$ is partial operation which can be defined in terms of the semiring's pre-order: $r \wedge s = t$ if $t \sqsubseteq r$, $t \sqsubseteq s$ and there exists no other $t'$ where $t' \sqsubseteq r$ and $t' \sqsubseteq s$ and $t \sqsubseteq t'$.} $\bigwedge^n_{i =
  1} r_i$ is the resulting grade (see
\granin{pullPair} from Section~\ref{sec:motivating-example}).

For types $A$ closed with respect to recursion variables, let $\deriv{pull}{A} = \lambda z
. \deriv{pull}{A}^\emptyset\ z$ where:
\begin{align*}
  \deriv{pull}{\mathbf{1}}^\Sigma            \ z & = \textbf{case} \  \GRANULEmv{z}  \ \textbf{of} \    \mathsf{unit}    \rightarrow  \GRANULEsym{[}    \mathsf{unit}    \GRANULEsym{]}
\\
\deriv{pull}{\alpha}^\Sigma      \ z & = z
                                       \\
\deriv{pull}{X}^\Sigma           \ z & = \Sigma(X)\ z
                                    \\
\deriv{pull}{A \oplus B}^\Sigma  \ z & =\synCaseTwo{z}{\mathsf{inl}\
                                       x}{\synCaseOne{\deriv{pull}{A}^{\Sigma}\ x}{[u]}{[\mathsf{inl}\ u]}}{\mathsf{inr}\ y}
                            {\synCaseOne{\deriv{pull}{B}^{\Sigma}\
                                       y}{[v]}{[\mathsf{inr}\ v]}}
\\
\deriv{pull}{A \otimes B}^\Sigma \ z & =
\synCaseOne{z}{(x, y)}
   {\synCaseOne{(\deriv{pull}{A}^{\Sigma}\ x, \deriv{pull}{B}^{\Sigma}\ y)}
                {([u], [v])}{[(u, v)]}} \\
\deriv{pull}{\mu X . A}^\Sigma \ z & =
   \synLetRec{f}{\deriv{pull}{A}^{\Sigma, X \mapsto f : \mu {X} .    \GRANULEnt{A}  \overrightarrow{ [    \square_{\textcolor{coeffectColor}{ \GRANULEmv{r_{\GRANULEmv{i}}} } }  \alpha_{\GRANULEmv{i}}    /  \alpha_{\GRANULEmv{i}}   ] }      \multimap   \square_{\textcolor{coeffectColor}{   \bigwedge_{i = 1}^n  \GRANULEmv{r_{\GRANULEmv{i}}}   } }  \GRANULEsym{(}   \mu {X} .  \GRANULEnt{A}   \GRANULEsym{)} }}{f\ z}
 \end{align*}
%
%
Just like \emph{push}, we cannot apply \emph{pull} to graded modalities themselves. Unlike
\emph{push}, we cannot apply \emph{pull} to function types. That is, we
cannot derive a distributive law of the form $(\Box_r A \multimap \Box_r B)
\multimap \Box_r (A \multimap B)$ since introducing the concluding $\Box_r$
would require the incoming function $(\Box_r A \multimap \Box_r B)$ to
itself be inside $\Box_r$ due to the promotion rule (\textsc{pr}), which does not
match the type scheme for \emph{pull}.

The rest of the derivation above is
similar but dual to that of \emph{push}.

The appendix~\cite{appendix} proves that
$\deriv{pull}{A}$ is type sound, i.e., its derivations are well-typed.

\begin{example}
  \label{ex:push-fun}
  To illustrate the above procedures, the derivation of
  $\deriv{push}{(\alpha \otimes \alpha) \multimap \beta}$ is:
  \begin{align*}
       & \lambda z . \deriv{push}{(\alpha \otimes \alpha) \multimap \beta}^\emptyset \ z
         : \Box_r ((\alpha \otimes \alpha) \multimap \beta) \multimap
             ((\Box_r \alpha \otimes \Box_r \alpha) \multimap \Box_r \beta)
\\
  = \; & \lambda z . \lambda y . \synCaseOne{z}{[f]}{\synCaseOne{\deriv{pull}{\alpha
      \otimes \alpha}^{\emptyset}\
         y}{[u]}{\deriv{push}{\beta}^{\emptyset}[(f \ u)]}}
    \\
  = \; & \lambda z . \lambda y . \synCaseOne{z}{[f]}{\synCaseOne{(\synCaseOne{y}{(x', y')}
   {\\ & \hspace{11em} \synCaseOne{(\deriv{pull}{\alpha}^{\emptyset}\ x', \deriv{pull}{\alpha}^{\emptyset}\ y')}
                {([u], [v])}{[(u, v)]}})}{[u]}{\deriv{push}{\beta}^{\emptyset}[(f \ u)]}}
    \\
  = \; & \lambda z . \lambda y . \synCaseOne{z}{[f]}{\synCaseOne{(\synCaseOne{y}{(x', y')}
   {\synCaseOne{(x', y')}
                {([u], [v])}{[(u, v)]}})}{[u]}{[(f \ u)]}}
  \end{align*}
\end{example}

\begin{remark}
One might ponder whether linear logic's exponential $!
A$~\cite{girard1987linear} is modelled by the graded necessity modality over
$\mathbb{N}_{\infty}$ intervals, i.e., with $! A \triangleq \Box_{\textcolor{coeffectColor}{{  \textcolor{coeffectColor}{0}  }..{  \infty  } }} A$. This is a reasonable assumption, but
$\Box_{\textcolor{coeffectColor}{{  \textcolor{coeffectColor}{0}  }..{  \infty  } }} A$ has a slightly different meaning to $! A$,
exposed here: whilst $\deriv{push}{A \otimes B} : \Box_{\textcolor{coeffectColor}{{  \textcolor{coeffectColor}{0}  }..{  \infty  }}} (A \otimes B) \multimap (\Box_{\textcolor{coeffectColor}{{  \textcolor{coeffectColor}{0}  }..{  \infty  } }} A \otimes \Box_{\textcolor{coeffectColor}{{  \textcolor{coeffectColor}{0}  }..{  \infty  }}} B)$
is derivable in \grminip{}, linear logic does not permit $!(A \otimes B) \multimap
(!A \otimes !B)$. Models of $!$ provide only a monoidal functor
structure which gives \emph{pull} for $\otimes$, but not
\emph{push}~\cite{benton1992linear}. This structure can be recovered in Granule through
the introduction of a partial type-level operation which selectively
disallows \emph{push} for $\otimes$ in semirings which model the $!$
modality of linear logic\footnote{The work in ~\cite{hughes:lirmm-03271465} arose as a result of the writing of this paper.} ~\cite{hughes:lirmm-03271465}.
\end{remark}

The algorithmic definitions of `push' and `pull' can be leveraged
in a programming context to automatically yield these combinators for
practical purposes. We discuss how this is leveraged inside the
Granule compiler in Section~\ref{sec:implementation} and two techniques
for leveraging it for Linear Haskell in Section~\ref{sec:linhaskell}.
Before that, we study the algebraic behaviour of the derived distributive laws.

\subsection{Properties}
\label{subsection:properties}


We consider here the properties of these derived operations.  Prima
facie, the above \emph{push} and \emph{pull} operations are simply
distributive laws between two (parametric) type constructors $\tcF$
and $\Box_r$, the latter being the graded modality. However, both
$\tcF$ and $\Box_r$ have additional structure. If the mathematical
terminology of `distributive laws' is warranted, then such additional structure
should be preserved by \emph{push} and \emph{pull} (e.g., as in how
a distributive law between a monad and a comonad must preserve
the behaviour of the monad and comonad operations after applying
the distributive law~\cite{power2002combining}); we explain here the relevant
additional structure and verify the distributive law properties.

Firstly, we note that these distributive laws are mutually inverse:

\begin{restatable}[Pull is right inverse to push]{prop}{pushPullInverse}
  For all $n$-arity types $\tcF$ which do not contain function types,
  then for type variables $(\alpha_i)_{i \in [1..n]}$
  and for all grades $r \in \mathcal{R}$ where $\textcolor{coeffectColor}{1}   \, \textcolor{coeffectColor}{\sqsubseteq} \,  \GRANULEmv{r}$ if $|\tcF
    \overline{\alpha_i}| > 1$, then:
  \begin{align*}
\deriv{pull}{\tcF\
  \overline{\alpha_i}}(\deriv{push}{\tcF\
  \overline{\alpha_i}}) = id\ : \Box_{r} \tcF \overline{\alpha_i}
\multimap \Box_{r} \tcF \overline{\alpha_i}
    \end{align*}
\end{restatable}

\begin{restatable}[Pull is left inverse to push]{prop}{pullPushInverse}
  For all $n$-arity types $\tcF$ which do not contain function types,
  then for type variables $(\alpha_i)_{i \in [1..n]}$
  and for all grades $r \in \mathcal{R}$ where $\textcolor{coeffectColor}{1}   \, \textcolor{coeffectColor}{\sqsubseteq} \,  \GRANULEmv{r}$ if $|\tcF
    \overline{\alpha_i}| > 1$, then:
  \begin{align*}
\deriv{push}{\tcF\
  \overline{\alpha_i}}(\deriv{pull}{\tcF\
  \overline{\alpha_i}}) = id\ : \tcF (\Box_{r} \overline{\alpha_i})
\multimap \tcF (\Box_{r} \overline{\alpha_i})
    \end{align*}
  \end{restatable}

\noindent
The appendix~\cite{appendix} gives the proofs, leveraging the
equational theory of \grminip{}.

Applying a mathematical perspective, $\Box_r$ is also an endofunctor with
its object mapping provided by the type constructor itself and its
morphism mapping behaviour defined as follows:
\begin{definition}[$\Box_r$ functor]
  Given a function $f : \alpha \multimap \beta$ (a closed function
  term) then $\Box_r f : \Box_r \alpha \multimap \Box_r \beta$ is the
  morphism mapping of the endofunctor $\Box_r$ defined:
  \begin{align*}
    \Box_r\ f = \lambda x. \textbf{case} \  \GRANULEmv{x}  \ \textbf{of} \  \GRANULEsym{[}  \GRANULEmv{y}  \GRANULEsym{]}  \rightarrow  \GRANULEsym{[}  \GRANULEmv{f} \, \GRANULEmv{y}  \GRANULEsym{]}
  \end{align*}
\end{definition}
\noindent
For types $\tcF \alpha$ we can also automatically derive the
morphism mapping of a covariant functor, which we write as $\deriv{fmap}{\tcF \alpha}$
whose definition is standard (e.g., applied in Haskell~\cite{generic-deriving})  given in
the appendix~\cite{appendix}. Distributive laws between
endofunctors should be natural transformations, which is indeed the case for our
derivations:

\begin{restatable}[Naturality of push]{prop}{pushNatural}
  For all unary type constructors $\tcF$ such that $\deriv{push}{\tcF \alpha}$ is defined, and given a closed function term $f : \alpha \multimap \beta$, then: $ {\deriv{fmap}{\tcF} \Box_{r}f} \circ \deriv{push}{\tcF {\alpha}} = \deriv{push}{\tcF {\beta}} \circ  \Box_{r} \deriv{fmap}{\tcF} f $, i.e.:
\begin{align*}
\xymatrix@C=3.5em{
{\alpha} \ar[d]_{f}
\\
{\beta} &
}
\quad
\xymatrix@C=5em{
{\Box_{r} \tcF {{\alpha}}} \ar[d]_{\Box_{r} \deriv{fmap}{\tcF } f } \ar[r]^{\deriv{push}{\tcF {\alpha}}}  &   {\tcF \Box_{r} {\alpha}}
\ar[d]^{\deriv{fmap}{\tcF } \Box_{r}f}   \\
{\Box_{r} \tcF {{\beta}}}  \ar[r]_{\deriv{push}{\tcF {\beta}}} &  {\tcF \Box_{r} {\beta}}}
\end{align*}
\end{restatable}

\begin{restatable}[Naturality of pull]{prop}{pullNatural}
  For all unary type constructors $\tcF$ such that $\deriv{pull}{\tcF \alpha}$ is defined, and given a closed function term $f : \alpha \multimap \beta$, then: $\Box_{r} \deriv{fmap}{\tcF} f \circ \deriv{pull}{\tcF {\alpha}} = \deriv{pull}{\tcF {\beta}} \circ {\deriv{fmap}{\tcF} \Box_{r}f}$, i.e.:
\begin{align*}
\xymatrix@C=3.5em{
{\alpha} \ar[d]_{f}
\\
{\beta} &
}
\quad
\xymatrix@C=5em{
{\tcF \Box_{r} {\alpha}} \ar[d]_{\deriv{fmap}{\tcF}\Box_{r}f} \ar[r]^{\deriv{pull}{\tcF {\alpha}}}  & {\Box_{r} \tcF {{\alpha}}}
\ar[d]^{ \Box_{r}\deriv{fmap}{\tcF} f}   \\
{ \tcF \Box_{r} {\beta}}  \ar[r]_{\deriv{pull}{\tcF {\beta}}} &  {\Box_{r} \tcF {{\beta}}}}
\end{align*}
\end{restatable}
\noindent
The appendix~\cite{appendix} gives the proofs.
Note that the naturality results here are for cases of unary type
constructors $\tcF$ that are covariant functors, written with a single parameter $\alpha$. This can
easily generalise to $n$-ary type constructors.

Not only is $\Box_r$ an endofunctor but it also has the structure
of a \emph{graded comonad}~\cite{combining2016,DBLP:conf/fossacs/Katsumata18,petricek2014coeffects,DBLP:conf/icalp/PetricekOM13}.

\begin{definition}[Graded comonadic operations]
The \grminip{} calculus (and Granule) permits the derivation of graded
comonadic
operations~\cite{DBLP:journals/pacmpl/OrchardLE19}
for the semiring graded necessity $\Box_r$, defined:
\begin{align*}
\varepsilon_A & : \Box_1 A \multimap A = \lambda x . \textbf{case} \  \GRANULEmv{x}  \ \textbf{of} \  \GRANULEsym{[}  \GRANULEmv{z}  \GRANULEsym{]}  \rightarrow  \GRANULEmv{z} \\
\delta_{r,s,A} & : \Box_{r \ast{} s} A \multimap \Box_r \Box_s A
= \lambda x . \textbf{case} \  \GRANULEmv{x}  \ \textbf{of} \  \GRANULEsym{[}  \GRANULEmv{z}  \GRANULEsym{]}  \rightarrow  \GRANULEsym{[}  \GRANULEsym{[}  \GRANULEmv{z}  \GRANULEsym{]}  \GRANULEsym{]}
 \end{align*}
 \end{definition}
\noindent
 The derived distributive laws preserve these graded comonadic
 operations i.e., the distributive laws are
 well-behaved with respect to the graded comonadic structure of
 $\Box_r$, captured by the following properties:

\begin{restatable}[Push preserves graded comonads]{prop}{pushPreserve}
For all $\tcF$ such that $\deriv{push}{\tcF \overline{\alpha_i}}$ is defined
and $\tcF$ does not contain $\multimap$ (to avoid issues of
contravariance in $\tcF$) then:
\begin{align*}
\xymatrix@C=3.5em@R=1.4em{
\Box_{1} \tcF \overline{\alpha_i} \ar[r]^{\deriv{push}{\tcF
  \overline{\alpha_i}}} \ar[d]_{\varepsilon}
& \tcF \overline{\Box_{1} \alpha_i}
\ar[dl]^{\tcF \varepsilon}
\\
\tcF \overline{\alpha_i} &
}
\quad
\xymatrix@C=5em@R=1.4em{
\Box_{r\ast{}s}
\tcF \overline{\alpha_i} \ar[d]_{\delta_{r,s}} \ar[rr]^{\deriv{push}{\tcF \overline{\alpha_i}}}
& & \tcF \overline{\Box_{r\ast{}s} \alpha_i} \ar[d]^{\tcF \delta_{r,s}}
\\
\Box_{r} \Box_{s} \tcF \overline{\alpha_i} \ar[r]_{\Box_r \deriv{push}{\tcF \overline{\alpha_i}}}&
\Box_{r}\tcF \overline{\Box_{s} \alpha_i} \ar[r]_{\deriv{push}{\tcF \overline{\alpha_i}}} &
\tcF \overline{\Box_{r} \Box_{s} \alpha_i}
&
}
\end{align*}
\end{restatable}

\begin{restatable}[Pull preserves graded comonads]{prop}{pullPreserve}
\label{prop:pullPreserve}
%
%
For all $\tcF$ such that $\deriv{pull}{\tcF \overline{\alpha_i}}$ is defined then:
\begin{align*}
\xymatrix@C=3.5em@R=1.4em{
\Box_{1} \tcF \overline{\alpha_i} \ar[d]_{\varepsilon}
&  \ar[l]_{\deriv{pull}{\tcF \overline{\alpha_i}}} \tcF
  \overline{\Box_{1} \alpha_i}
\ar[dl]^{\tcF \varepsilon}
\\
\tcF \overline{\alpha_i} &
}
\quad
\xymatrix@C=5em@R=1.4em{
\Box_{r\ast{}s}
\tcF \overline{\alpha_i} \ar[d]_{\delta_{r,s}}
& & \ar[ll]_{\deriv{pull}{\tcF \overline{\alpha_i}}} \tcF
    \overline{\Box_{r\ast{}s} \alpha_i}
\ar[d]^{\tcF \delta_{r,s}}
\\
\Box_{r} \Box_{s} \tcF \overline{\alpha_i} & \ar[l]^{\Box_r \deriv{pull}{\tcF \overline{\alpha_i}}}
\Box_{r}\tcF \overline{\Box_{s} \alpha_i} & \ar[l]^{\deriv{pull}{\tcF \overline{\alpha_i}}}
\tcF \overline{\Box_{r} \Box_{s} \alpha_i}
&
}
\end{align*}
\end{restatable}
\noindent
The appendix~\cite{appendix} gives the proofs.

\section{Implementation in Granule}
\label{sec:implementation}
The Granule type checker implements the algorithmic derivation of \emph{push}
and \emph{pull} distributive laws as covered in the previous section. Whilst the
syntax of \grminip{} types had unit, sum, and product types as primitives, in
Granule these are provided by a more general notion of type constructor which
can be extended by user-defined, generalized algebraic data types (GADTs). The
procedure outlined in Section~\ref{sec:push-pull} is therefore generalised
slightly so that it can be applied to any data type: the case for $A \oplus B$
is generalised to types with an arbitrary number of data constructors.

Our deriving mechanism is exposed to programmers via explicit (visible) type
application (akin to that provided in GHC Haskell~\cite{eisenberg2016visible})
on reserved names \granin{push} and \granin{pull}. Written \granin{push @T} or
\granin{pull @T}, this signals to the compiler that we wish to derive the
corresponding distributive laws at the type \granin{T}. For example, for the
\granin{List : Type -> Type} data type from the standard library, we can write
the expression \granin{push @List} which the type checker recognises as a
function of type:
\begin{granule}
push @List : forall {a : Type, s : Semiring, r : s} . {1 <= r} => (List a) [r] -> List (a [r])
\end{granule}
Note this function is not only polymorphic in the grade, but polymorphic in the
semiring itself. Granule identifies different graded modalities by
their semirings, and thus this operation is polymorphic in the graded
modality. When the type checker encounters such a type application, it
triggers the derivation procedure of Section~\ref{sec:push-pull},
which also calculates the type. The result is then stored in the state
of the frontend to be passed to the interpreter (or compiler) after
type checking. The derived operations are memoized so that they need not be re-calculated if a particular distributive law is required more than once.
Otherwise, the implementation largely follows
Section~\ref{sec:push-pull} without surprises,
apart from some additional machinery for specialising the types
of data constructors coming from (generalized) algebraic data types.

\paragraph{Examples}
Section~\ref{sec:intro} motivated the crux of this paper with a
concrete example, which we can replay here in concrete Granule, using
its type application technique for triggering the automatic derivation
of the distributive laws. Previously, we defined \granin{pushPair} by
hand which can now be replaced with:
\begin{granule}
push @(,) : forall {a, b : Type, s : Semiring, r : s} . (a, b) [r] -> (a [r], b [r])
\end{granule}
Note that in Granule \granin{(,)} is an infix type constructor for products as well as terms. We could then replace the previous definition of \granin{fst'} from
Section~\ref{sec:intro} with:
\begin{granule}
fst' : forall {a, b : Type, r : Semiring} . {0 <= r} => (a, b) [r] -> a
fst' = let [x'] = fst (push @(,) x) in x'
\end{granule}
The point however in the example is that we need not even define this
intermediate combinator, but can instead write the following
wherever we need to compute the first projection
of \granin{myPair : (a, b) [r]}:
\begin{granule}
extract (fst (push @(,) myPair)) : a
\end{granule}
We already saw that we can then generalise this by applying
this first projection inside of the list \\ \granin{myPairList : (List (a, b))
  [r]} directly, using \granin{push @List}.

In a slightly more elaborate example, we can use the \granin{pull} combinator
for pairs to implement a function that duplicates a pair (given that both elements
can be consumed twice):
\begin{granule}
copyPair : forall {a, b : Type} . (a [0..2], b [2..4]) -> ((a, b), (a, b))
copyPair x = copy (pull @(,) x) -- where, copy : a [2] -> (a, a)
\end{granule}
Note \granin{pull} here computes the greated-lower
bound of intervals \granin{0..2} and \granin{2..4} which is
\granin{2..2}, i.e., we can provide a pair of \granin{a} and \granin{b}
values which can each be used exactly twice, which is what is required
for \granin{copy}.

As another example, interacting with Granule's indexed types
(GADTs), consider a simple programming task of taking the head of a sized-list (vector)
and duplicating it into a pair. The \granin{head} operation
is typed: \granin{head : forall \{a : Type, n : Nat\} . (Vec (n + 1) a)
  [0..1] -> a} which has a graded modal input with grade \granin{0..1} meaning
the input vector is used 0 or 1 times:
the head element is used once (linearly) for the return
but the tail is discarded.

This head element can then be copied
if it has this capability via a graded modality, e.g., a value of type \granin{(Vec (n + 1) (a [2]))
  [0..1]} permits:
\begin{granule}
copyHead' : forall {a : Type, n : Nat} . (Vec (n + 1) (a [2])) [0..1] -> (a, a)
copyHead' xs = let [y] = head xs in (y, y) -- [y] unboxes (a [2]) to y:a usable twice
\end{granule}
Here we ``unbox'' the graded modal value of type \granin{a [2]} to
get a non-linear variable \granin{y} which we can use precisely twice.
However, what if we are in a programming
context where we have a value \granin{Vec (n + 1) a} with no
graded modality on the type \granin{a}? 
We can employ two idioms here:
(i) take a value of type \granin{(Vec (n + 1) a) [0..2]} and
split its modality in two: \granin{(Vec (n + 1) a) [2] [0..1]}
(ii) then use \textit{push} on the inner graded modality
\granin{[2]} to get
\granin{(Vec (n + 1) (a [2])) [0..1]}.

Using \granin{push @Vec} we can thus write the following to duplicate
the head element of a vector:
\begin{granule}
copyHead : forall {a : Type, n : Nat} . (Vec (n + 1) a) [0..2] -> (a, a)
copyHead = copy . head . boxmap [push @Vec] . disject
\end{granule}
which employs combinators from the standard library and
the derived distributive law, of type:
\begin{granule}
boxmap    : forall {a b : Type, s : Semiring, r : s}        . (a  -> b) [r] -> a [r] -> b [r]
disject   : forall {a : Type, s : Semiring, n m : s}        . a [m * n] -> (a [n]) [m]
push @Vec : forall {a : Type, n : Nat, s : Semiring, r : s} . (Vec n a) [r] -> Vec n (a [r])
\end{granule}

\section{Application to Linear Haskell}
\label{sec:linhaskell}
\lstset{
  language=Haskell,
  basicstyle=\ttfamily\footnotesize,
  literate={\\\%}{\%}2
}

While Granule has been pushing the state-of-the-art in graded modal types,
similar features have been added to more mainstream languages. Haskell has
recently added support for linear types via an underlying
graded system which enables linear types as a smooth extension
to GHC's current type system~\cite{linear-haskell}.\footnote{Released
as part of GHC v9.0.1 in February 2021 \url{https://www.haskell.org/ghc/download_ghc_9_0_1.html}}
Functions may be linear with respect to
their input (meaning, the function will consume its argument exactly
once if the result is consumed exactly once), but
they can also consume their argument \granin{r}
times for some multiplicity \granin{r} via explicit `multiplicity polymorphism'.
Unlike Granule, Linear Haskell limits this multiplicity to the set of either
one or many (the paper speculates about extending this with zero)---full natural
numbers or other semirings are not supported.

In Linear Haskell, the function type (\haskin{a \%r -> b})
can be read as ``a function from \granin{a}
to \granin{b} which uses \granin{a} according to \granin{r}'' where
\granin{r} is either $1$ (also written as \haskin{'One}) or $\omega$
(written as \haskin{'Many}).
For example, the following defines and types the linear and non-linear functions
\granin{swap} and \granin{copy} in Linear Haskell:
\begin{haskell}
{-# LANGUAGE LinearTypes #-}
import GHC.Types

swap :: (a %1 -> b %1 -> c) %1 -> (b %1 -> a %1 -> c)
swap f x y = f y x

copy :: a %'Many -> (a, a)
copy x = (x, x)
\end{haskell}
Assigning the type \haskin{a \%1 -> (a, a)} to
\haskin{copy} would result in a type error due to a mismatch in
multiplicity.

The approach of Linear Haskell (as formally defined by Bernardy et al.~\cite{linear-haskell})
resembles earlier coeffect
systems~\cite{ghica2014bounded,petricek2014coeffects} and
more recent work on graded systems which have appeared
since~\cite{abel-barnardy-icfp2020,choudhury2021}. In
these approaches there is no underlying linear type system (as there is in Granule) but
grading is instead pervasive with function arrows carrying a grade
describing the use of their parameter. Nevertheless, recent systems also provide a graded modality
as this enables more fine-grained usage information to be ascribed to compound data.
For example, without graded modalities it cannot be explained that the first projection on
a pair uses the first component once and its second component not at
all (instead a single grade would have to be assigned to the entire pair).

We can define a graded modality in Linear Haskell
via the following \haskin{Box} data type that is parameterized
over the multiplicity \granin{r} and the value type \granin{a}, defined as:
\begin{haskell}
data Box r a where { Box :: a %r -> Box r a }
\end{haskell}
A \granin{Box} type is necessary
to make explicit the notion that a value may be consumed a certain number of times,
where ordinarily Linear Haskell is concerned only with whether individual functions
consume their arguments linearly or not.
Thus, distributive laws of the form discussed in this paper
then become useful in practice when working with Linear Haskell.

The \granin{pushPair} and \granin{pullPair} functions from Section~\ref{sec:intro}
can then be implemented in Linear Haskell with the help of this \granin{Box} type:
\begin{haskell}
pushPair :: Box r (a, b) %1 -> (Box r a, Box r b)
pushPair (Box (x, y)) = (Box x, Box y)
\end{haskell}
\begin{haskell}
pullPair :: (Box r a, Box r b) %1 -> Box r (a, b)
pullPair (Box x, Box y) = Box (x, y)
\end{haskell}
Interestingly, \haskin{pushPair} could also be implemented as a function of type
\haskin{(a, b) \%r -> (Box r a, Box r b)}, and in general we can formulate
push combinators typed \haskin{(f a) \%r -> f (Box r a)}, i.e., consuming
the input \granin{r} times, returning a box with multiplicity \granin{r}, but
we stick with the above formulation for consistency.

While more sophisticated methods are outlined in this paper for automatically
synthesizing these functions in the context of Granule, in the
context of Linear Haskell (which has a simpler notion of grading) distributive laws over unary and binary type
constructors can be captured with type classes:
\begin{haskell}
class Pushable f where
  push :: Box r (f a) %1-> f (Box r a)
class Pushable2 f where
  push2 :: Box r (f a b) %1-> f (Box r a) (Box r b)

class Pullable f where
   pull :: f (Box r a) %1-> Box r (f a)
class Pullable2 f where
  pull2 :: f (Box r a) (Box r b) %1-> Box r (f a b)
\end{haskell}
Separate classes here are defined for unary and binary cases, as working generically
over both is tricky in Haskell.
Implementing an instance of \granin{Pushable2} for pairs is then:
\begin{lstlisting}[language=Haskell]
instance Pushable2 (,) where
  push2 (Box (x, y)) = (Box x, Box y)
\end{lstlisting}
This implementation follows the procedure of
Section~\ref{sec:push-pull}.
A pair type $A \otimes B$ is handled by pattern matching on \granin{(Box (x, y))} and boxing both
fields in the pair \granin{(Box x, Box y)}.

A Haskell programmer may define instances of these type classes
for their own data types, but as discussed in Section~\ref{sec:intro}, this is
tedious from a software engineering perspective.
Template Haskell is a meta-programming system for Haskell that allows
compile-time generation of code, and one of the use cases for Template Haskell
is generating boilerplate code~\cite{template-haskell}. The instances of
\granin{Pushable} and \granin{Pullable} for algebraic data types are relatively
straightforward, so we implemented procedures that will automatically generate
these type class instances for arbitrary user-defined types (though types with
so-called ``phantom'' parameters are currently not supported).

For example, if a programmer wanted to define a \granin{Pushable} instance for
the data type of a linear \granin{List a}, they would write:
\begin{lstlisting}[language=Haskell,mathescape=true]
data List a where
  Cons :: a %1-> List a %1-> List a
  Nil  :: List a
$\$$(derivePushable ''List)
\end{lstlisting}
Here, \granin{derivePushable} is a Template Haskell procedure\footnote{Available online: \url{https://github.com/granule-project/deriving-distributed-linear-haskell}} that takes a name
of a user-defined data type and emits a top-level declaration, following the
strategy outlined in Section~\ref{sec:push-pull}.
For the \granin{List a} data type above, we can walk through the type as
\granin{derivePushable} would. Because \granin{List a} has two constructors, a case statement
is necessary (our code generator will use a `case lambda'). This case will have branches for each
of the \granin{Cons} and \granin{Nil} constructors---in the body of the former it must box the
first field (of type \granin{a}) and recursively apply \granin{push} to the second field (of
type \granin{List a}) after boxing it, and in the body of the latter it must simply return
\granin{Nil.}
The full code generated by \granin{derivePushable ''List} is given below.
\begin{lstlisting}
derivePushable ''List
======>
instance Pushable List where
  push
    = \case
        Box (Cons match_a4BD match_a4BE)
          -> (Cons (Box match_a4BD)) (push (Box match_a4BE))
        Box Nil -> Nil
\end{lstlisting}
Later, in Section~\ref{sec:other}, we will discuss other combinators and type classes that are
useful in Linear Haskell.

The applicability of our proposed deriving method to Haskell shows that it is
useful beyond the context of the Granule project. Despite Haskell not having a formal semantics, we believe the same equational reasoning can be applied to show that the properties in Section~\ref{subsection:properties} also hold for these distributive laws in Linear Haskell. As more programming languages
adopt type system features similar to Granule and Linear Haskell, we expect that
deriving/synthesizing or otherwise generically implementing combinators derived
from distributive laws will be increasingly useful.







\section{Typed-analysis of Consumption in Pattern Matching}
\label{sec:matching-and-consumption}
\label{subsec:matching-and-consumption}

\newcommand{\abname}{$\Lambda^p$}
\newcommand{\gradname}{\textsc{GraD}}

This paper's study of distributive laws provides
an opportunity to consider design decisions for the \emph{typed
  analysis of pattern matching} since the operations of
Section~\ref{sec:push-pull} are derived by pattern matching in concert
with grading. We compare here the choices made surrounding the typing of
pattern matching in four works (1) Granule and
its core calculus~\cite{DBLP:journals/pacmpl/OrchardLE19} (2) the graded
modal calculus \abname{} of Abel and Bernardy~\cite{abel-barnardy-icfp2020} (3) the dependent graded system
\gradname{} of Choudhury et al.~\cite{choudhury2021} and (4) Linear
Haskell~\cite{linear-haskell}.

\paragraph{Granule}
Pattern matching against a graded modality $\Box_r A$ (with pattern
$[p]$) in Granule is provided by the \textsc{Pbox} rule
(Figure~\ref{fig:pattern-rules}) which triggers typing pattern $p$
`under' a grade $r$ at type $A$. This was denoted via the optional grade information
$\GRANULEmv{r}   \vdash \,  \GRANULEnt{p}  :  \GRANULEnt{A}  \, \rhd \,  \Gamma$ which then pushes grading down onto the
variables bound within $p$. Furthermore, it is only under such a
pattern that wildcard patterns are allowed (\textsc{[Pwild]}),
requiring $\coeff{\textcolor{coeffectColor}{0}   \, \textcolor{coeffectColor}{\sqsubseteq} \,  \GRANULEmv{r}}$, i.e., $\GRANULEmv{r}$ can approximate
$\textcolor{coeffectColor}{0}$ (where $\coeff{0}$ denotes weakening).
None of the other systems considered here have such a facility
for weakening via pattern matching.

For a type $A$ with more than one
  constructor ($|  \GRANULEnt{A}  | > 1$), pattern matching its
  constructors underneath an $r$-graded box requires $\textcolor{coeffectColor}{1}   \, \textcolor{coeffectColor}{\sqsubseteq} \,  \GRANULEmv{r}$. For example, eliminating sums inside an $\coeff{\GRANULEmv{r}}$-graded box
$\square_{\textcolor{coeffectColor}{ \GRANULEmv{r} } }  \GRANULEsym{(}   \GRANULEnt{A}  \, \oplus \,  \GRANULEnt{B}   \GRANULEsym{)}$ requires $\textcolor{coeffectColor}{1}   \, \textcolor{coeffectColor}{\sqsubseteq} \,  \GRANULEmv{r}$ as
distinguishing $\mathsf{inl}$ or $\mathsf{inr}$
constitutes a \emph{consumption} which reveals information (i.e.,
pattern matching on the `tag' of the data constructors).
By contrast, a type with only one constructor cannot convey any information
by its constructor and so matching on it is not counted as a consumption:
eliminating $\square_{\textcolor{coeffectColor}{ \GRANULEmv{r} } }  \GRANULEsym{(}   \GRANULEnt{A}  \, \otimes \,  \GRANULEnt{B}   \GRANULEsym{)}$ places no
requirements on $r$.
The idea that unary data types do not incur consumption (since no
information is conveyed by its constructor) is a refinement here to the
original Granule paper as described by Orchard et
al.~\cite{DBLP:journals/pacmpl/OrchardLE19}, which for \textsc{[Pcon]}
had only the premise $\textcolor{coeffectColor}{1}   \, \textcolor{coeffectColor}{\sqsubseteq} \,  \GRANULEmv{r}$ rather than
$|  \GRANULEnt{A}  | > 1 \implies \textcolor{coeffectColor}{1}   \, \textcolor{coeffectColor}{\sqsubseteq} \,  \GRANULEmv{r}$ here (although the
implementation already reflected this idea).

\paragraph{The \abname{} calculus}
Abel and Bernardy's unified modal system \abname{} is akin to Granule,
but with pervasive grading (rather than base linearity) akin to the
coeffect calculus~\cite{petricek2014coeffects} and Linear
Haskell~\cite{linear-haskell} (discussed in
Section~\ref{sec:linhaskell}). Similarly to the situation in Granule,
\abname{} also places a grading requirement when pattern matching on a
sum type, given by the following typing rule in their syntax~\cite[Fig
1, p.4]{abel-barnardy-icfp2020}:
\begin{align*}
\dfrac{\gamma \Gamma \vdash t : A_1 + A_2 \qquad \delta\Gamma, x_i :^q
  A_i \vdash u_i : C \qquad q \leq 1}
      {(q\gamma + \delta)\Gamma \vdash \mathsf{case}^q\ t\ \mathsf{of}\
  \{\mathsf{inj}_1 x_1 \mapsto u_1; \mathsf{inj}_2 x_2 \mapsto u_2 \}
  : C}\textsc{$+$-elim}
\end{align*}
The key aspects here are that variables $x_i$ bound in the case are used
with grade $q$ as denoted by the graded assumption $x_i :^q A_i$ in the context of
typing $u_i$ and then that $q \leq 1$ which is exactly our
constraint that $\textcolor{coeffectColor}{1}   \, \textcolor{coeffectColor}{\sqsubseteq} \,  \GRANULEmv{r}$ (their ordering just runs in the
opposite direction to ours). For the elimination of pair and unit
types in \abname{} there is no such constraint, matching our idea that arity
affects usage, captured in Granule by $|  \GRANULEnt{A}  | > 1 \implies \textcolor{coeffectColor}{1}   \, \textcolor{coeffectColor}{\sqsubseteq} \,  \GRANULEmv{r}$. Their typed-analysis of patterns is motivated
by their parametricity theorems.

\paragraph{\gradname{}}
The dependent graded type system \gradname{} of Choudhury et al. also considers
the question of how to give the typing of pattern matching on sum
types, with a rule in their system~\cite[p.8]{choudhury2021} which
closely resembles the \textsc{$+$-elim} rule for \abname{}:
\begin{align*}
\dfrac{
  \Delta ; \Gamma_1 \vdash q : A_1 \oplus A_2
  \qquad
  \Delta ; \Gamma_2 \vdash b_1 : A_1 \xrightarrow{q} B
  \qquad
  \Delta ; \Gamma_2 \vdash b_2 : A_2 \xrightarrow{q} B
  \qquad
  1 \leq q
}
{\Delta ; q \cdot \Gamma_1 + \Gamma_2 \vdash \mathbf{case}_q\ a\
  \mathbf{of}\ b_1; b_2 : B}
\textsc{STcase}
\end{align*}
%
The direction of the preordering in \gradname{} is the same
as that in Granule but, modulo this ordering and some slight
restructuring, the case rule
captures the same idea as \abname{}: ``both branches
of the base analysis \emph{must} use the scrutinee at least once,
as indicated by the $1 \leq q$ constraint.''~\cite[p.8]{choudhury2021}.
Choudhury et al., also provide a heap-based semantics which serves to connect the
meaning of grades with a concrete operational model of usage, which
then motivates the grading on sum elimination here. 
In the simply-typed version of \gradname{}, matching on the components
of a product requires that each component is consumed linearly.

\paragraph{Linear Haskell}
The paper on Linear Haskell by Bernardy et
al.~\cite{linear-haskell} has a \textbf{case} expression for
eliminating arbitrary data constructors, with grading similar
to the rules seen above. Initially, this rule is for the setting
of a semiring over $\mathcal{R} = \{1, \omega\}$ (described
in Section~\ref{sec:linhaskell}) and has no requirements on the grading
to represent the notion of inspection, consumption, or usage due to
matching on (multiple) constructors.
This is reflected in the current
implementation where we can define the following sum elimination:
\begin{haskell}
match :: (Either a b) %r -> (a %1 -> c) -> (b %1 -> c) -> c
match (Left x) f _  = f x
match (Right x) _ g = g x
\end{haskell}
However, later when considering
the generalisation to other semirings they state that
``\emph{typing rules are mostly unchanged with the caveat that
$\mathsf{case}_\pi$ must exclude $\pi = 0$}''~\cite[\S{}7.2,
p.25]{linear-haskell} where $\pi$ is the grade of the $\textbf{case}$ guard.
This appears a more coarse-grained restriction
than the other three systems, excluding even the possibility of
Granule's weakening wildcard pattern which requires $0 \leq \pi$.
Currently, such a pattern must be marked as \haskin{'Many} in Linear
Haskell (i.e., it cannot explain that first projection on a pair
does not use the pair's second component).
Furthermore, the condition $\pi \neq 0$ does not require that $\pi$
actually
represents a consumption, unlike the approaches of the other three
systems.
%
The argument put forward by Abel and Bernardy for their restriction to
mark a consumption ($q \leq 1$)
for the sake of parametricity is a compelling one, and the concrete model of
Choudhury et al. gives further confidence that this restriction captures well an
operational model.
Thus, it seems there is a chance for fertilisation
between the works mentioned here and Linear Haskell's vital work, towards a
future where grading is a key tool in the typed-functional programmer's toolbox.

\section{Deriving Other Useful Structural Combinators}
\label{sec:other}
So far we have motivated the use of distributive laws, and
demonstrated that they are useful in practice when programming in
languages with linear and graded modal types.
The same methodology we have been discussing can also be used to
derive other useful generic combinators for programming with linear
and graded modal types. In this section, we consider two structural
combinators, \granin{drop} and \granin{copyShape}, in Granule as well
as related type classes for dropping, copying, and moving resources in Linear Haskell.

\subsection{A Combinator for Weakening (``drop'')}
\label{subsec:drop}

%
The built-in type constants of Granule can be split into those
which permit structural weakening $C^{\mathsf{w}}$ such as
\granin{Int}, \granin{Char}, \granin{String}, and those which do not
$C^{\mathsf{l}}$ such as \granin{Handle} (file handles) and
\granin{Chan} (concurrent channels). Those that
permit weakening contain non-abstract values that
can in theory be systematically inspected in order to consume them.
Granule provides a built-in implementation of \granin{drop}
for $C^{\mathsf{w}}$ types, which is then used by the following derivation
procedure to derive weakening on compound types:
\begin{align*}
\deriv{drop}{A} : A \multimap 1
\end{align*}
for closed types $A$ defined
$\deriv{drop}{A} = \lambda z . \deriv{drop}{A}^\emptyset z$
by an intermediate derivation $\deriv{drop}{A}^\Sigma$:
\begin{align*}
\deriv{drop}{C^w}^\Sigma z & = \text{\granin{drop} $z$} \\
\deriv{drop}{1}^\Sigma z & = \textbf{case} \  \GRANULEmv{z}  \ \textbf{of} \    \mathsf{unit}    \rightarrow    \mathsf{unit} \\
\deriv{drop}{X}^\Sigma z & = \Sigma(X) z \\
\deriv{drop}{A \oplus B}^\Sigma z & =
\synCaseTwoShort{z}{\mathsf{inl}\ x}{\deriv{drop}{A}(x)}{\mathsf{inr}\ y}
                            {\deriv{drop}{B}(y)}\\
\deriv{drop}{A \otimes B}^\Sigma z & =
\synCaseOne{z}{(x, y)}
   {\synCaseOne{\deriv{drop}{A}(x)}{\mathsf{unit}}
     {\synCaseOne{\deriv{drop}{B}(y)}{\mathsf{unit}}{\mathsf{unit}}}}
\\
\deriv{drop}{\mu X . A}^\Sigma & z =
                                 \synLetRec{f}{\deriv{drop}{A}^{\Sigma,
                                 X \mapsto f : A \multimap 1}}{f\ z}
\end{align*}
Note we cannot use this procedure in a polymorphic context (over type
variables $\alpha$) since type polymorphism ranges over all types,
including those which cannot be dropped like $C^{\mathsf{l}}$.


\subsection{A Combinator for Copying ``shape''}


The ``shape'' of values for a parametric data types $\tcF$ can be determined
by a function $\mathit{shape} : \tcF A \rightarrow \tcF 1$, usually
derived when $\tcF$ is a functor by mapping with $A \rightarrow 1$
(dropping elements)~\cite{jay1994shapely}. This provides a way of
capturing the size, shape, and form of a data structure.
Often when programming with data structures which must be
used linearly, we may wish to reason about properties of the data structure
(such as the length or ``shape'' of the structure) but we may not
be able to drop the contained values. Instead, we wish to extract
the shape but without consuming the original data structure itself.

This can be accomplished with a function which copies
the data structure exactly, returning this duplicate along with a data
structure of the same shape, but with the
terminal nodes replaced with values of the unit type $1$  (the `spine'). For example, consider a
pair of integers: \granin{(1, 2)}. Then applying \granin{copyShape} to this pair
would yield \granin{(((), ()), (1, 2))}. The original input pair is duplicated
and returned on the right of the pair, while the left value contains
a pair with the same structure as the input, but with values replaced with
\granin{()}. This is useful, as it allows us to use the left value of the
resulting pair to reason about the structure of the input (e.g., its
depth / size), while preserving the original input. This is
particularly useful for deriving size and length combinators for
collection-like data structures.

\noindent
As with ``drop'', we can derive such a
function automatically:
\begin{align*}
\deriv{copyShape}{\tcF \alpha} : \tcF \alpha \multimap \tcF 1 \otimes \tcF \alpha
\end{align*}
defined by
$\deriv{copyShape}{A} = \lambda z . \deriv{copyShape}{A}^\emptyset z$
by an intermediate derivation $\deriv{copyShape}{A}^\Sigma$:
\begin{align*}
\deriv{copyShape}{C^w}^\Sigma z & = (\mathsf{unit},\ z) \\
\deriv{copyShape}{1}^\Sigma z & = \synCaseOne{z}{\mathsf{unit}}{(   \mathsf{unit}    ,    \mathsf{unit}   )}\\
\deriv{copyShape}{\alpha}^\Sigma z & = (\mathsf{unit}, z) \\
\deriv{copyShape}{X}^\Sigma z & = \Sigma(X) z \\
\deriv{copyShape}{A \oplus B}^\Sigma z & =
                                         \synCaseTwo{z}{\mathsf{inl}\ x}{
             \synCaseOne{\llbracket{   \GRANULEnt{A}   }\rrbracket_{\mathsf{copyShape} }^{ \Sigma }  \GRANULEsym{(}  \GRANULEmv{x}  \GRANULEsym{)}}{ ( \GRANULEmv{s},\ \GRANULEmv{x'} )}{(   \mathsf{inl}\  \GRANULEmv{s}    ,    \mathsf{inl}\  \GRANULEmv{x'}   )}} {\mathsf{inr}\ y}
                            {
             \synCaseOne{\llbracket{   \GRANULEnt{B}   }\rrbracket_{\mathsf{copyShape} }^{ \Sigma }  \GRANULEsym{(}  \GRANULEmv{y}  \GRANULEsym{)}}{ ( \GRANULEmv{s},\ \GRANULEmv{y'} )}{(   \mathsf{inr}\  \GRANULEmv{s}    ,    \mathsf{inr}\  \GRANULEmv{y'}   )}}\\
\deriv{copyShape}{A \otimes B}^\Sigma z & =
\synCaseOne{z}{(x, y)}
   {
                                          \synCaseOne{\llbracket{   \GRANULEnt{A}   }\rrbracket_{\mathsf{copyShape} }^{ \Sigma }  \GRANULEsym{(}  \GRANULEmv{x}  \GRANULEsym{)}}{ ( \GRANULEmv{s},\ \GRANULEmv{x'} )}{
                                          \\ & \qquad  \synCaseOne{\llbracket{   \GRANULEnt{B}   }\rrbracket_{\mathsf{copyShape} }^{ \Sigma }  \GRANULEsym{(}  \GRANULEmv{y}  \GRANULEsym{)}}{ ( \GRANULEmv{s'},\ \GRANULEmv{y'} )}{
                                          ((\GRANULEmv{s},\ \GRANULEmv{s'}),\ (\GRANULEmv{x'},\ \GRANULEmv{y'}))}
                                          }}
\\
\deriv{copyShape}{\mu X . A}^\Sigma & z =
                                 \synLetRec{f}{\deriv{copyShape}{A}^{\Sigma,
                                 X \mapsto f : A \multimap 1 \otimes A}}{f\ z}
\end{align*}
The implementation recursively follows the structure of the type,
replicating the constructors, reaching the crucial case where a
polymorphically type $z : \alpha$ is mapped to $(   \mathsf{unit}    ,  \GRANULEmv{z} )$ in
the third equation.

Granule implements both these derived combinators in a similar
way to \emph{push}/\emph{pull} providing \granin{copyShape}
and \granin{drop} which can be derived for a type \granin{T} via type
application, e.g. \granin{drop @T : T -> ()} if it can be derived.
Otherwise, the type checker produces an error,
explaining why \granin{drop} is not derivable at type \granin{T}.

\subsection{Other Combinators in Linear Haskell}




As previously covered in Section~\ref{sec:linhaskell}, we demonstrated that the
\emph{push} and \emph{pull} combinators derived from distributed laws are useful
in Haskell with its linear types extension, and we demonstrated that they can be
automatically derived using compile-time meta-programming with Template Haskell.

To the best of our knowledge, nothing comparable to the \granin{Pushable} and
\granin{Pullable} type classes proposed here has been previously discussed in
the literature on Linear Haskell. However, several other type classes have been
proposed for inclusion in the standard library to deal with common use cases
when programming with linear function types.\footnote{See the
\texttt{linear-base} library: \url{https://github.com/tweag/linear-base}.} One
of these classes, \haskin{Consumable}, roughly corresponds to the
\granin{drop} combinator above, while the other two, \haskin{Dupable} and
\haskin{Movable}, are for when a programmer wants to allow a data type to be
duplicated or moved in linear code.
\begin{lstlisting}[language=Haskell]
class Consumable a where
  consume :: a %1-> ()

class Consumable a => Dupable a where
  dup2 :: a %1-> (a, a)

class Dupable a => Movable a where
  move :: a %1-> Ur a
\end{lstlisting}
The \haskin{consume} function is a linear function from a value to unit,
whereas \haskin{dup} is a linear function from a value to a pair of that
same value. The \haskin{move} linear function maps \haskin{a} to
\haskin{Ur a}, where \haskin{Ur} is the ``unrestricted'' modality.
Thus, \granin{move} can be used to implement both \granin{consume} and \granin{dup2}:
\begin{lstlisting}[language=Haskell]
case move x of {Ur _ -> ()}     -- consume x
case move x of {Ur x -> x}      -- x
case move x of {Ur x -> (x, x)} -- dup2 x
\end{lstlisting}
A `copy shape' class may also be a useful addition to Linear
Haskell in the future.


\section{Discussion and Conclusion}
\label{sec:conclusion}

Section~\ref{sec:matching-and-consumption} considered, in some detail,
systems related to Granule and different typed analyses
of pattern matching. Furthermore, in
applying our approach to both Granule and Linear Haskell
we have already provided some detailed comparison between the two. This section
considers some wider related work alongside ideas for future work,
then concludes the paper.

\paragraph{Generic Programming Methodology}
The deriving mechanism for Granule is based on the methodology of
generic functional programming~\cite{hinze2000new}, where functions
may be defined generically for all possible data types in the language;
generic functions are defined inductively on the structure of the types.
This technique has notably been used before in Haskell, where there
has been a strong interest in deriving type
class instances automatically. Particularly relevant to this paper is the work
on generic deriving~\cite{generic-deriving}, which allows Haskell programmers to
automatically derive arbitrary class instances using standard datatype-generic
programming techniques as described above. In this paper we opted to rely on
compile-time metaprogramming using Template Haskell~\cite{template-haskell}
instead, but it is possible that some of the combinators we describe could be
implemented using generic deriving as well, which is future work.

\paragraph{Non-graded Distributive Laws}
Distributive laws are standard components in abstract
mathematics. Distributive laws between categorical structures used for
modelling modalities (like monads and comonads) are well explored. For
example, Brookes and Geva defined a categorical semantics using monads
combined with comonads via a distributive law capturing both
intensional and effectful aspects of a
program~\cite{brookes1993intensional}. Power and Watanabe study in
detail different ways of combining comonads and monads via
distributive laws~\cite{power2002combining}. Such distributive laws
have been applied in the programming languages literature, e.g., for
modelling streams of partial elements~\cite{uustalu2006essence}.

\paragraph{Graded Distributive Laws}
Gaboardi et al. define families of graded distributive laws
for graded monads and comonads~\cite{combining2016}. They
include the ability to interact the grades, e.g., with operations
such as $\Box_{\iota(r,f)} \Diamond_f A \rightarrow \Diamond_{\kappa(r,f)} \Box_r A$
between a graded comonad $\Box_r$ and graded monad $\Diamond_f$ where
$\iota$ and $\kappa$ capture information about the distributive law
in the grades. In comparison, our distributive laws here are more
prosaic since they involve only a graded comonad (semiring graded
necessity) distributed over a functor and vice versa. That said,
the scheme of Gaboardi et al. suggests that there might be interesting
graded distributive laws between $\Box_r$ and the indexed types,
for example, $\Box_r (\mathsf{Vec}\, n \, A) \rightarrow \mathsf{Vec}\,
  (r * n) \, (\Box_1 A)$ which internally replicates a
  vector. However, it is less clear how useful such combinators would
be in general or how systematic their construction would be. In
contrast, the distributive laws explained here appear frequently
and have a straightforward uniform calculation.

We noted in Section~\ref{sec:push-pull} that neither of our
distributive laws can be derived over graded modalities themselves,
i.e., we cannot derive
$\textit{push} : \Box_r \Box_s A \rightarrow \Box_s \Box_r A$. Such an
operation would itself be a distributive law between two graded modalities,
which may have further semantic and analysis consequences beyond the normal derivations here
for regular types. Exploring this is future work, for which the previous
work on graded distributive laws can provide a useful
scheme for considering the possibilities here.
Furthermore, Granule has both graded comonads and graded monads
so there is scope for exploring possible graded distributive laws
between these in the future following Gaboardi et
al.~\cite{combining2016}.

In work that is somewhat adjacent to this paper, we define program
synthesis procedures for graded linear
types~\cite{DBLP:conf/lopstr/HughesO20}. This program synthesis
approach can synthesis \emph{pull} distributive laws that are
equivalent to the algorithmically derived operations of this
paper. Interestingly the program synthesis approach cannot derive the
\emph{push} laws though as its core theory has a simplified notion of
pattern typing that doesn't capture the full power of Granule's
pattern matches as described in this paper.

\paragraph{Conclusions}
The slogan of graded types is to imbue types with
information reflecting and capturing the underlying program semantics
and structure. This provides a mechanism for fine-grained intensional
reasoning about programs at the type level, advancing the power of
type-based verification. Graded types are a burgeoning technique that is
gaining traction in mainstream functional programming and is being
explored from multiple different angles in the literature. The
work described here addresses the practical aspects of applying
these techniques in real-world programming. Our hope is that this
aids the development of the next generation of programming languages
with rich type systems for high-assurance programming.

\paragraph{Acknowledgments} Thanks to Harley Eades III and Daniel Marshall
for discussion and comments on a draft and to the reviewers and
participants at TLLA-Linearity 2020.  This work was supported by the
EPSRC, grant EP/T013516/1 \emph{Verifying Resource-like Data Use in
  Programs via Types}. The first author is also supported by an EPSRC
DTA.

\nocite{*}
\bibliography{references}

\ifappendix
\newpage
\appendix

\clearpage
\pagenumbering{arabic}
\setcounter{page}{1}
\thispagestyle{empty}
\section*{Deriving Distributive Laws for Graded Linear Types (Additional Material)}
{\large{Jack Hughes, Michael Vollmer, Dominic Orchard}}\\
{{School of Computing, University of Kent}}\\\\
This document contains the supplementary material for the paper of the same
name which appears in the EPTCS proceedings for TLLA/Linearity 2020. Included are the full proofs for all propositions which appear in the paper, as well as any additional definitions that are required.
\input{proofs-ottput}
\fi

\end{document}